\documentclass[apj]{emulateapj}
\usepackage{natbib}
\usepackage{ifpdf}
\slugcomment{Accepted for publication in ApJ}

\usepackage{graphicx}
\usepackage{amssymb}
\usepackage{subfigure}
\newcommand{\msun}{~\mbox{M}_{\odot}}

\newcommand\zsun{{\rm \,Z_\odot}}
\newcommand\lsun{{\rm \,L_\odot}}
\newcommand{\unit}[1]{\ensuremath{\, \mathrm{#1}}}

\newcommand{\cMpc}{~\mbox{comoving}~\mbox{Mpc}}

\newcommand{\cmci}{~\mbox{cm}^{-3}}

\newcommand{\Msun}{~\mbox{M}_{\odot}}

\begin{document}

\shorttitle{Simulating the fossils of the first galaxies}
\shortauthors{JEON ET AL.}

\title{Connecting the first galaxies with ultra faint dwarfs in the Local Group: chemical signatures of Population~III stars}
\author{Myoungwon Jeon\altaffilmark{1,3}, 
            Gurtina Besla\altaffilmark{1}
            Volker Bromm\altaffilmark{2}
            }
            
\altaffiltext{1}{Department of Astronomy, University of Arizona, 933 North Cherry Avenue, Tucson, AZ 85721, USA; myjeon@email.arizona.edu}
\altaffiltext{2}{Department of Astronomy, University of Texas, TX 78712, USA}
\altaffiltext{3}{Center for Global Converging Humanities, Kyung Hee University, Republic of Korea}

\begin{abstract}
We investigate the star formation history and chemical evolution of isolated analogues of Local Group (LG) 
ultra faint dwarf galaxies (UFDs; stellar mass range of $10^2\Msun<M_{\ast}<10^5\Msun$) and gas rich, 
low mass dwarfs (Leo P analogs; stellar mass range of $10^5\Msun<M_{\ast}<10^6\Msun$). We perform a suite of 
cosmological hydrodynamic zoom-in simulations to follow their evolution
from the era of the first generation of stars down to $z=0$. 
We confirm that reionization, combined with 
supernova (SN) feedback, is primarily responsible for the truncated star formation in UFDs.
Specifically, haloes with a virial mass of 
$\rm M_{\rm vir}\lesssim2\times10^9\Msun$ form $\gtrsim90\%$ of stars prior to reionization. 
Our work further demonstrates the importance of Population~III (Pop~III) stars, with their intrinsically high $\rm [C/Fe]$ yields, and the associated 
external metal-enrichment, in producing low-metallicity stars ($\rm [Fe/H]\lesssim-4$) and carbon-enhanced 
metal-poor (CEMP) stars. We find that UFDs are composite systems, assembled from multiple progenitor haloes, 
some of which hosted only Population~II (Pop~II) stars formed in environments externally enriched
by SNe in neighboring haloes, naturally producing, extremely low-metallicity Pop~II stars. 
We illustrate how the simulated chemical enrichment may be used to constrain the star formation histories (SFHs) of true observed UFDs. 
We find that Leo P analogs can form in haloes with $\rm M_{\rm vir} \sim 4\times 10^9 \Msun$ ($z=0$). 
Such systems are less affected by reionization and continue to form stars until z=0, causing higher metallicity tails. 
Finally, we predict the existence of extremely low-metallicity stars in LG UFD galaxies 
that preserve the pure chemical signatures of Pop~III nucleosynthesis. 

%These predictions can be probed with 
%upcoming deep observations of metal-poor stars in the LG. 
%We show that $\alpha$-elements are enhanced at all metallicities, in particular for haloes with $\rm M_{\rm vir}\lesssim2\times10^9\Msun$, 
%implying that the gas in UFDs has preferentially been contaminated by Type~II SNe from Pop~III and massive Pop~II stars. 
%promising to offer crucial 
%insights into the nature of the first stars and the chemical 
%enrichment history in the early Universe. 
\end{abstract}


\section{Introduction}
\label{Sec:Intro}

One of the biggest challenges in modern cosmology is to understand the first generation of 
stars and galaxies that formed during the cosmic Dark Ages (see, e.g. \citealp{Bromm2011} for a review). 
They reside in observationally unexplored territory, rendering their detection one of the main goals 
of the upcoming next-generation facilities, such as the {\it James Webb Space Telescope (JWST)} 
(e.g. \citealp{Gardner2006}) and the 30-40~m giant ground-based telescopes.
%Recent theoretical studies suggest that the first galaxies were metal-poor and small in size, with 
%a stellar mass of $M_{\ast}\lesssim10^6\Msun$ ($z\gtrsim7$). These properties are 
%similar to the faintest dwarf galaxies in the Local Group (LG)  (e.g. \citealp{Greif2009}; \citealp{Wise2012};
% \citealp{Jeon2015}; \citealp{Ricotti2016}). 
The lowest mass ($10^2\Msun\lesssim M_{\ast}\lesssim10^5\Msun$, $z=0$) and most metal-poor ($\rm [Fe/H]\lesssim-2$) dwarfs of the Local 
Group (LG), known as Ultra Faint Dwarfs (UFDs), have been proposed as possible descendants of the dwarf galaxies at high redshift and are thus the 
best place to look for the metal poor stars that hold clues to the nature of the first generation of stars.
Studying such systems enables an approach often termed ``galactic archaeology", wherein local galaxies are used as fossil 
records of the early assembly history of the Milky Way (MW) 
(e.g. \citealp{Salvadori2009}; \citealp{Frebel2010}; \citealp{Boylan-Kolchin2015}; \citealp{Ricotti2016}; \citealp{Lapi2017}; \citealp{Weisz2017}). 
%GB adding a sentence to fit your work 
Here we illustrate how high resolution cosmological simulations that follow structure formation in the smallest haloes at 
high redshift can help link the first galaxies to their local 
descendants and identify the chemical signatures of the first generation of stars, the so-called Population~III (Pop~III) stars, in the 
stellar populations of present day UFDs and low-mass dwarfs.
\par

The dwarf galaxy satellites of the MW exhibit a great diversity in stellar age, metallicity, and sizes, indicating that they have experienced 
different star formation histories (SFHs). Specifically, more massive dwarfs ($M_{\ast} > 10^6\Msun$, $z=0$)  
are characterized by
extended SFHs and show large metallicity spreads, whereas the lower mass UFDs appear to have experienced a truncated SFH, yielding
uniformly ancient stellar populations, older than $\sim$11-12 Gyr (e.g. \citealp{Bullock2000}; \citealp{Grebel2004}; \citealp{Brown2012}; 
 \citealp{Weisz2014}; \citealp{Brown2014}). 
\par
A plausible mechanism responsible for such early suppression of star formation in UFDs could be reionization, 
referring to the global phase transition from a neutral Universe to an ionized one at $z=10-6$ (\citealp{planck2015}). 
During this period of time, the radiation from stars heated the 
gas in the interstellar medium (ISM) and intergalactic medium (IGM), thus boosting the Jeans mass,
 i.e., the minimum mass to trigger gravitational instability, which 
in turn made it difficult for gas to collapse into haloes (see, e.g. \citealp{Tolstoy2009} for a review; and 
\citealp{Gnedin2006}; \citealp{Bovill2009}; \citealp{Bovill2011}; \citealp{Brown2012}). 

A new puzzle has recently been presented in the discovery of low mass, gas-rich dwarfs in the outskirts of the Milky Way
(e.g. \citealp{McQuinn2015}). 
Dwarfs like Leo P, Leo T, Leo A and DDO 210 have stellar masses less than $10^6 \Msun$ ($z=0$), but are able to retain substantial 
gas reservoirs ($\rm M_{\rm HI} \sim 10^{5-7} \Msun$, $z=0$). 
%To date a cosmological study for the origin of such systems has not been conducted. 
Here, we explore for the first time the connection between these gas rich, low mass dwarfs 
and present-day UFDs using cosmological zoom-in simulations that follow their evolution from the first generation of stars to z=0. 

Significant theoretical progress has been made towards understanding the formation of low mass dwarf galaxies 
(e.g. \citealp{Salvadori2009}; \citealp{Sawala2011}; \citealp{Munshi2013}; \citealp{Simpson2013}; 
\citealp{Governato2015}; \citealp{Wheeler2015}; \citealp{Onorbe2015}; \citealp{Fitts2016}). 
However, the emphasis of these existing studies has been to understand the SFHs of dwarfs and their structural properties at $z=0$. In particular, work by \citet{Wheeler2015} and \citet{Simpson2013} used cosmological zoom simulations of the formation 
of low mass galaxies ($M_{\ast}\lesssim10^6\Msun$ at $z=0$) at high redshift to demonstrate that a combination 
of stellar feedback and reionization can result in the suppression of star formation in low mass haloes ($M_{\ast} \lesssim 10^6\msun$ at $z=0$). 
However, these works only followed the chemical evolution of total metallicity, rather than individual metal species, and 
none tracked the metal-enrichment history from Pop~III stars to low redshift. Furthermore, existing work concerning the chemical abundances is typically considered in  
an idealized setting (e.g. \citealp{Revaz2009}; \citealp{Corlies2013}; \citealp{Webster2014}; \citealp{Webster2015}; \citealp{Bland-Hawthorn2015}; \citealp{Revaz2016}).
As such, the connection of the stellar populations of LG UFDs and low-mass dwarfs to those of the first generation of stars remains uncertain. 
We seek to address this major gap in knowledge by self-consistently following the metal enrichment and diffusion from SNe  
from the first generation of stars in these low mass galaxies, 
accounting for both reionization and their subsequent hierarchical evolution until the present day.

\citet{Ricotti2005} classify dwarf galaxies based on their SFHs into three categories: galaxies in which the bulk of stars were formed prior to reionization are 
classified as ``true fossils", whereas ``polluted fossils" refer to galaxies that have continuously formed stars beyond reionization, and  
``survivors" mean galaxies that formed the majority of their stars after reionization.  By following the evolution of the 
first galaxies until the present, we will be able to establish the expected degree of ``pollution" in UFDs and furthermore 
quantify the significance of late time accretion events to their present day chemical properties. 
%% This allows you to later discuss expected differences between the UFDs that were never within the virial radius 
%% of the MW and those that were accreted at late times.  (or in other words, ways to identify their accretion epoch**) 
%% Uma would be expected to have been accreted a bit later than the others - something that might be revealed by 
%% its orbital properties (HST on going programs) .  -- Need to add a discussion section about this. 
True fossils, in particular, could offer more stringent insight on the nature of the 
first generation of stars, preserving their chemical signatures since these 
galaxies are not likely to undergo significant star formation after reionization.
\par

The SN feedback from Pop~III stars drove the gas out of minihaloes, 
thus polluting the surrounding IGM with the metals (e.g. \citealp{Kitayama2005}; \citealp{Ritter2012}; \citealp{Jeon2014}). 
The second generation, call Population~II (Pop~II) stars, were born out of the gas thus shaped by Pop~III SNe, and, 
as a result, they might preserve the chemical signatures of the Pop~III stars. Unlike Pop~III stars, the lower mass of Pop~II stars can allow them to survive until today. 
Therefore, investigating the chemical abundance patterns of Pop~II stars may inform us of the star formation at early times in 
UFDs and low-mass dwarfs. In particular, the abundances of alpha elements compared to iron can tell us about 
the timescale over which stars have formed and about the corresponding star formation efficiency. 
%For example, 
%alpha-elements such as Si, Mg, Ca, and Ti are more abundant relative to iron in Type~II SNe from massive stars, 
%whereas iron dominates over alpha elements when Type~Ia SNe are triggered (\citealp{Tinsley1979}). 
%Owing to the long lifetime of low-mass stars, the onset of Type~Ia SNe is delayed, on the order of $\sim$ 100~Myr or longer
%after star formation begins. Consequently, the enhanced $[\alpha/\rm Fe]$ ratios from Type~II SNe decline with 
%increasing $\rm [Fe/H]$, as stars begin to form in gas clouds that are metal-enriched by Type~Ia SNe 
%after $\gtrsim$ 100 Myr. 
%This trend is also applicable to star formation in the early Universe. 
%The mass range of Pop~III stars is still uncertain, but 
%recent statistical studies of first star formation have predicted massive stars with a broad range from a few tens to a 
%few 1000$\msun$ (e.g. \citealp{Hirano2014}; \citealp{Hirano2015}). As a consequence, 
%Based on the expected mass range of Pop~III stars varying from a few $10\msun$ to $1000\msun$, they were likely to die as Type~II SNe rather 
%than Type~Ia, leading to an enhancement in [$\alpha/\rm Fe$] ratios (e.g. \citealp{Hirano2014}; \citealp{Hirano2015}). 
%These chemical peculiarities are expected to be preserved in the 
%subsequent generation of stars. Reionization additionally ``froze in'' any such signatures: if further star formation 
%was terminated by reionization in high-$z$ dwarfs, their SFHs would extend over $\sim$500-600 Myr, which is too short for
%Type~Ia SNe to become the dominant source of metals. 

To date, a complete theoretical understanding of chemical abundances in UFD stellar populations, 
in a full cosmological context, and considering the
detailed enrichment from early Pop~III and Pop~II stars, is still lacking. 
This study is the first to self-consistently follow these processes from high redshift to the present day, allowing us to 
constrain the expected signatures of Pop~III stars in the stellar populations of current UFDs and low-mass dwarfs.
%Note that the work presented here is broadly consistent with the  
%upcoming study by Corlies et al. (2017, priv. comm.), although their simulations stop shortly after reionization.

\par
Here, we present the results of a suite of cosmological, high-resolution zoom-in 
simulations that trace the chemical and mass evolution of low mass dwarf galaxies (UFD mass analogs) from high redshifts down to $z=0$.
In this work, we have two main goals: (1) to investigate the role of reionization and SNe feedback in the SFHs of UFD and low-mass, gas-rich dwarf analogs; and 
(2) to compare the resulting stellar abundance patterns with observational data from present day UFDs, highlighting signatures of chemical enrichment from Pop~III stars. 
We will discuss the structural properties of the simulated dwarfs (e.g. velocity dispersions and sizes) in an upcoming paper. 
%  compare physical quantities derived from the simulations, such as 
%stellar mass, luminosity, and stellar abundance pattern, with observational data.
%Specifically, for the first objective, we focus on the role of heating mechanisms, responsible for the suppression of star formation, such as
%SN feedback and reionization, and their effectiveness depending on halo mass.
\par 
The most important difference of this work from 
previous studies is the ab-initio inclusion of early star formation and enrichment. 
In addition, for the first time, we derive detailed chemical abundance patterns
 for individual stars from realistic cosmological simulations, to be compared with
stellar archaeological data.
Our hydrodynamical simulations are nicely complemented by the recent dark matter-only
investigation of \citet{Griffen2016}. Both treatments employ cosmological
initial conditions, with the latter focusing on representing the realistic
large-scale density field of the LG environment, while our work
includes the detailed physics of the baryonic component. 
Corlies et al. 2017 (priv. comm.) consider the SFHs and metallicity distribution 
functions of a range of halos within a fully simulated cosmological box as opposed 
to zooming in on specific halos. However, the computationally intensive feedback 
methods implemented restrict the simulation to a small volume and can only be run to z=7.  

\par
The outline of the paper is as follows. Our numerical methodology is described in Section~2, and the simulation results are presented 
in Section~3. We discuss the limitations of this work in Section~4. Finally, our main findings are summarized in Section~5.
For consistency, all distances are expressed in physical (proper) units unless noted otherwise.

\section{Numerical methodology}
In this Section, we present numerical methods we adopt for the simulations. At first, we explain initial conditions, followed 
by the cosmological and hydrodynamical parameters in Section~2.1. The implemented chemistry, cooling, and UV background are presented in Section~2.2. 
In Section~2.3, we describe star formation recipe for Pop~III and Pop~II stars. The accompanied chemical feedback of SNe from both 
populations is discussed in Section~2.4. Finally, we show how we include thermal feedback from SNe in Section~2.5. Note that 
possible caveats and limitations of the methods described here are pointed out in Section~4.

\begin{deluxetable*}{c c c c c c c c c c c c c}
\centering
\tablecolumns{11}
\tablewidth{0.85\textwidth}
\tablecaption{Characteristics of the simulated UFD and gas-rich dwarf analogs at $z=0$.}
\tablehead{ 
\colhead{Halo} & \colhead{$M_{\rm vir}$} & \colhead{$r_{\rm v}$} & \colhead{$M_{\ast}$} & \colhead{$D_{\rm h}$}  & \colhead{$f_{\rm b}$} & \colhead{$r_{1/2}^{\ast}$} & \colhead{$[\rm Fe/H]$} & \colhead{$\bar{[{\alpha/\rm Fe]}}$} & \colhead{$\sigma_{\ast}$} & \colhead{$M_{\rm gas}$} &  \colhead{$M_{\rm HI,w}$/$M_{\rm HI,c}$} & \colhead{$\rm SF_{\rm trun}$ }\\
\colhead{Unit} & \colhead{$[10^9\msun]$} & \colhead{$[\rm kpc]$} & \colhead{$[10^4\msun]$} &  \colhead{[Mpc]} & \colhead{[$\%$]}  & \colhead{$[\rm pc]$} & \colhead{-}  & \colhead{-} & \colhead{$[\rm km s^{-1}]$} &  \colhead{$[10^6\msun]$} &  \colhead{$[10^5\msun]$} &\colhead{-}
}
\startdata
\vspace{0.07cm}
halo1 & 1.53 & 23.7& 4.3&0.6& 0.08&345&-2.63&0.52&6.4 &1.30& - &Yes\\
\vspace{0.07cm}
halo2 & 1.53 & 23.5 & 3.8 & 2.0& 0.07&320 &-2.25&0.44&6.0& 1.15& - &  Yes\\
\vspace{0.07cm}
halo3 & 1.60 &23.9&8.2& 2.1&0.1&296&-2.28&0.52&6.7& 1.67& - & Yes\\
\vspace{0.07cm}
halo4 & 2.21 & 26.6 & 13.0& 1.9& 0.96& 513 &-2.45&0.54&11.2& 1.13& - & No\\
\vspace{0.07cm}
halo5 & 3.15 & 29.9 & 20.0& 0.9& 0.05&479 &-2.27&0.53&9.9&  1.58& - & No\\
\vspace{0.07cm}
halo6 & 3.95 &32.1&88.6& 3.7 &0.9&438&-1.23&0.47&11.6& 26&$44$/$0.59$& No
%\vspace{0.07cm}
%halo6 (z=3) & 0.65 & 7.4&58& n/a & 1.0 &302&-1.76& 0.49 &7.4& 26&$0$/$0$& No
\enddata
%\vspace{0.8cm}
\tablecomments{First row denotes the derived physical quantities of the galaxies. First column
indicates the name of haloes at each zoom-in region. Column (2): viral mass in $10^9\msun$.
Column (3): virial radius in kpc. Column (4): stellar mass in $10^4\msun$.
Column (5): distance from a MW like host halo in Mpc. Column (6): baryon fraction in $\%$. Column (7): half stellar mass radius in pc.
Column (8): average iron-to-hydrogen ratios of stars. Column (9): average alpha-to-iron ratios of stars. Column (10): stellar velocity
dispersion in km $\rm s^{-1}$. Column(11): total gas mass in $10^6\msun$. Column(12): warm and cold neutral hydrogen gas mass in $10^5\msun$, 
defined as $T<5000$ K and $n_{\rm H}>$ 0.4 $\rm cm^{-3}$ for the warm H~I gas, $M_{\rm HI,w}$, and $T<1000$ K and $n_{\rm H}>$ 10 $\rm cm^{-3}$ 
for the cold H~I gas, $M_{\rm HI,c}$, respectively. Column (13): truncated star formation after reionization.}
\end{deluxetable*}

\label{Sec:Metho}
\subsection{Simulation Set Up}
We have performed a suite of hydrodynamic zoom-in simulations using a modified version of the $N$-body/TreePM 
Smoothed Particle Hydrodynamics (SPH) code {\sc GADGET} (\citealp{Springel2001}; \citealp{Springel2005}). 
As cosmological parameters, we adopt a matter density of $\Omega_{\rm m}=1-\Omega_{\Lambda}=0.265$, 
baryon density of $\Omega_{\rm b}=0.0448$, present-day Hubble expansion rate of $H_0 = 71\unit{km\, s^{-1} Mpc^{-1}}$, 
a spectral index $n_{\rm s}=0.963$, and a normalization $\sigma_8=0.801$ (\citealp{Komatsu2011}), consistent with 
the most recent constraints from the {\it Planck} satellite {\citealp{planck2015}}. The initial 
conditions are generated using the cosmological initial conditions code {\sc MUSIC} (\citealp{Hahn2011}). 
As a preliminary run, we perform a dark matter only using $128^3$ particles in a $L=6.25 h^{-1} \cMpc$ box. 
Then, we identify six galaxies with a mass of $M_{\rm vir}\sim10^9\msun$ at $z=0$ around a 
MW-size halo ($M_{\rm vir}\sim2\times10^{12}\msun$) (see, Figure~1). We confirm that the selected dwarfs have 
maintained an isolated position, outside of the virial radius of the MW-size host halo throughout their evolution. We 
conduct four consecutive refinements for all particles within $\sim2R_{\rm vir}$ of the selected 
low mass haloes at $z=0$, where $R_{\rm vir}$ is a virial radius, as marked in Figure~1, defined as the radius at which the mass density is 200 times higher 
than the average density of the Universe. The resultant effective resolution is $2048^3$, giving rise to dark matter (DM) and gas masses in the most refined 
region of $m_{\rm DM} \approx 2000\msun$ and $m_{\rm SPH} \approx 495 \msun$, respectively.

The properties of the selected haloes (UFD analogs) at z=0 are listed in Table~1 and discussed in detail in Section~3. 
% in the different zoom-in 
%regions are listed in Table~1. 
% GB Added:  
% Add Halo 6 properties at z=3 to Table 1 
The most massive halo, {\sc Halo 6}, is our prime candidate analog of gas-rich, low-mass dwarfs
like Leo P,  Leo A, Leo T and DDO 210. 
%Note that its properties at z=3 (listed in Table~1) are reasonably comparable to those of local UFDs. This suggests 
%that present-day, gas-poor UFDs with metal rich tails may have been born in more massive dwarf halos that were quenched after capture by a massive host prior to z=3.
This will be discussed in more detail in Section~3.2.  

\par
We find that in order to minimize computational cost, we preferentially choose haloes in relatively isolated regions, 
distributed from 0.6 Mpc to 2.1 Mpc from the center of the MW-size halo, which is illustrated as the largest 
white circle in the left panel of Figure~1. This choice 
naturally allows us to exclude other possible processes that might affect the SFHs and chemical properties of the 
dwarfs, such as tidal 
interactions and ram-pressure stripping. We fix the softening lengths for DM and star 
particles as $\epsilon_{\rm DM}=$ 40 pc at all redshifts. We use adaptive softening length for the gas particles, 
where the softening length is proportional to the SPH kernel length with a minimum value of $\epsilon_{\rm gas, min}=2.8$ pc.
\par
\subsection{Chemistry, Cooling, and UV Background}
We solve the coupled, non-equilibrium rate equations every time-step for the primordial chemistry from nine atomic and molecular species 
($\rm H, H^{+}, H^{-}, H_{2}, H^{+}_2 , He, He^{+}, He^{++},$ and $\rm e^{-}$), as well as from the three
deuterium species $\rm D, D^{+}$, and HD (\citealp{Glover2007}) in the presence of the cosmic UV/X-ray background provided by \citet{Haardt2012}. 
The UV background is implemented in terms of ionization and heating rates of H~I, He~I and He~II as a function of redshift. 
Such background is an important factor in influencing the SFHs, in that the early presence of UV radiation is likely to 
suppress star formation early on. Here, we begin to include a UV background at $z=7$, and linearly increase its strength until $z=6$. At lower redshifts, we
incorporate the background at full amplitude, assuming that
reionization is complete at $z=6$ (e.g. \citealp{Gunn1965}; \citealp{Fan2007}). 
\par
Self-shielding 
of the dense gas is included such that the UV background is attenuated as a function of $\exp{(-N_{\rm H~I} \bar{\sigma}_{\rm ion}})$, 
where $N_{\rm H~I}=x n_{\rm H~I}$. Here, $x=h$ is the SPH kernel size, $n_{\rm H~I}$ the neutral hydrogen number density, 
and $\bar{\sigma}_{\rm ion}$ the frequency-averaged photoionization cross-section for H~I. In addition, the photodissociation
of molecular hydrogen, $\rm H_2$, by the UV radiation in the Lyman-Werner (LW) band 
(11.2 eV$-$13.6 eV) is considered (e.g. \citealp{Abel1997}), with a rate $k_{\rm LW}=1.38\times 10^{-12} {\rm s^{-1}} J_{21}$. Here, the
normalized LW mean intensity is $J_{\rm 21}=\bar{J}_{\nu}/(10^{-21} {\rm erg Hz^{-1} s^{-1} cm^{-2} sr^{-1}})$, with $\bar{J}_{\nu}(z)$ being 
the average mean intensity in the LW band, calculated from the spectra in \citet{Haardt2011}. The treatment for photodissociation of deuterated hydrogen HD is 
identical to that of $\rm H_2$. 
\par
We consider all relevant primordial 
cooling processes such as H and He collisional ionization, excitation and recombination cooling, bremsstrahlung, 
inverse Compton cooling, and collisional excitation cooling of $\rm H_2$ and HD. Additionally, gas cooling by 
metal species such as carbon, oxygen, silicon, magnesium, neon, nitrogen, and iron is taken into account under equilibrium 
conditions. Specifically, we employ the method implemented 
by \citet{Wiersma2009a}, where the cooling rates are computed element-by-element from pre-computed tables based on the photo-ionization 
code {\sc CLOUDY} (\citealp{Ferland1998}), by interpolating the value as a function of density, temperature, and metallicity. 
Note that we ignore $\rm H_2$ formation via dust and also dust cooling for the following reasons: (1) at high redshifts, $z>7$, the 
amount of dust is not sufficient to be a dominant source of $\rm H_2$ formation, particularly in small systems ($M_{\ast}\lesssim10^6\msun$);
and (2) at low redshifts, metal cooling dominates over $\rm H_2$ cooling.

\subsection{Star Formation Physics}

The first stars are
predicted to form out of primordial gas inside minihaloes ($M_{\rm vir}\sim10^5-10^6\msun$)
at $z\gtrsim15$ (see, e.g. \citealp{Bromm2013} for a review; \citealp{Haiman1996}; \citealp{Tegmark1997}).
%The fate of Pop~III stars
%and their chemical signatures depends on their initial mass. Stars in the mass range ($m_{\ast, \rm Pop~III}=10-40\msun$) ended
%their lives as core-collapse supernovae (CCSNe) or pair-instability supernovae (PISNe) with a progenitor mass of
%$m_{\ast, \rm Pop~III}=140-260\msun$ (e.g. \citealp{Heger2002}). The Pop~III initial mass function (IMF) is still a subject of 
debate (e.g. \citealp{Bromm2013}). The final mass of Pop~III stars is determined by the complex interplay between protostellar UV feedback,
gas accretion through a self-gravitating circumstellar disk, and its fragmentation via gravitational instability (e.g. \citealp{McKee2008}; \citealp{Hosokawa2011}; 
\citealp{Stacy2012}). Recent 2-D and 3-D radiation hydrodynamic 
simulations, including stellar feedback from a protostar, predict that the mass of Pop~III stars is broadly distributed from $m_{\ast}\lesssim1\msun$ to $m_{\ast}\sim1000\msun$ 
(e.g. \citealp{Hirano2014}; \citealp{Susa2014}; \citealp{Hosokawa2016}; \citealp{Stacy2016}). 
The statistical study by \citet{Hirano2015} suggests that the characteristic mass of Pop~III stars also sensitively depends on the photo-dissociation from external far- UV (FUV) radiation, 
emitted by neighboring Pop~III stars.
%: at $z\sim25$ the Pop~III mass is on the order of a few hundred solar mass since the star-forming primordial gas are exposed to 
%intense local FUV radiation from nearby Pop~III stars, which 
%leads to the elevated Jeans-mass of the clouds. However, as the local FUV radiation becomes weaker owing to, in part, the expansion of the Universe, 
%at $z\lesssim15$ the masses of Pop~III stars decrease to a few tens of solar mass.
\par
Due to the limited numerical resolution, individual stars cannot be resolved, and we instead allow Pop~III stars to form as a single star cluster. 
We assume a top-heavy IMF with a functional form of $\phi_{\rm Pop~III}(m)=dN/d\log m=$const., and a mass range of $[m_0, m_1]=[10, 150]\msun$. Once a gas particle exceeds a threshold density of $n_{\rm H, th}= 100 \cmci$, the highest-density SPH particle is converted into a collisionless star particle with a 100$\%$ conversion efficiency, 
forming $M_{\ast}\approx500\msun$ in Pop~III stars at once. We note that the adapted threshold density is somewhat lower than the characteristic value, $n_{\rm H, th}= 10^4 \cmci$, where gravitational instability is triggered in primordial star formation (e.g. \citealp{Bromm2002}; \citealp{Abel2002}). In an effort to reduce computational cost, we avoid implementing such a higher threshold density. However, our choice of $n_{\rm H, th}= 100 \cmci$ is reasonable given that, above $n_{\rm H}= 10^3 \cmci$, the Pop~III star formation time is insensitive to any further increase in threshold density (\citealp{Muratov2013}). 

Stars are formed from gas clouds at a rate $\dot{\rho}_{\ast}=\rho_{\rm th}/\tau_{\ast}$, where $\tau_{\ast}=\tau_{\rm ff}/\epsilon_{\rm ff}$ is the star formation time scale, $\tau_{\rm ff}=[3\pi/(32G\rho_{\rm th})]^{1/2}$ the free fall time at the threshold density $\rho_{\rm th}$, and $\epsilon_{\rm ff}$ the star formation efficiency per free fall time (\citealp{Schmidt1959}). The star formation efficiency for Pop~III stars has yet to be pinned down precisely (e.g. \citealp{Hirano2015}; \citealp{Stacy2016}). Using a global baryon fraction of $f_{\rm b}=0.168$, we set $\epsilon_{\rm ff, Pop~III}=M_{\rm char, Pop~III}/(M_{\rm vir}\times f_{\rm b})$, where $M_{\rm char, Pop~III}$ is the characteristic mass of Pop~III stars, leading to $\epsilon_{\rm ff, Pop~III}\sim0.01$. We note that this value is similar to a typical efficiency in local star formation. Then, an SPH particle is stochastically converted into a collisionless star particle in a time interval $\Delta t$, if a random number is smaller than min($\Delta t$/$\tau_{\ast}$,1), to follow a given distribution $\dot{\rho}_{\ast}=\rho_{\rm th}/\tau_{\ast}$. The star formation timescale is then given by
\begin{equation}
\tau_{\ast}=\frac{\tau_{\rm ff} (n_{\rm H, th})}{\epsilon_{\rm ff}}\sim400 {\rm Myr} \left(\frac{n_{\rm H, th}}{100 \cmci}\right)^{-1/2}.
\end{equation}

Once pristine gas is enriched with metals dispersed by SN explosions of the first generation of stars, low-mass, long-lived second generation
Pop~II stars are formed out of the metal-polluted gas clouds. For Pop~II star formation, we employ the same star formation recipe as for Pop~III,
with an identical star formation efficiency, $\epsilon_{\rm ff, Pop~II}=0.01$,
but adding an additional metallicity criterion. If the metallicity of a gas 
particle, eligible for star formation, exceeds the critical metallicity, $Z_{\rm crit}=10^{-5.5}\zsun$, for the transition from 
Pop~III to Pop~II, we form a Pop~II star cluster. The choice of $Z_{\rm crit}=10^{-5.5}\zsun$ is motivated by 
dust-continuum cooling (e.g. \citealp{Omukai2000}; \citealp{Schneider2010}; \citealp{Bromm2001a};  \citealp{SSC2016}), where 
dust cooling is responsible for further gas fragmentation at high densities, $n_{\rm H}\gtrsim10^{16}\rm cm^{-3}$, enabling the formation of low-mass stars. 
For the Pop~II IMF, we use a Chabrier IMF over the mass range of $[0.1-100]\msun$. 
%Note that we treat both 
%Pop~II and Population~I as the same populations, not distinguish them.
%The star formation efficiency 
%of Pop~II stars out of the metal-enriched clouds is uncertain. Recent study done by \citet{SSC2015} shows that metal-enriched atomic cooling halo forms a 
%stellar mass of $\sim300\msun$ out of clumps at the initial stage $\sim10 \unit{kyr}$ and predicts that the final stellar mass of 
%$3000\msun$ if the entire mass of pre-stellar clumps is converted into stars. Assuming that their results can be considered as a representative of 
%how stars form in all clumps, we forms $m_{\ast, \rm Pop~II}=3000\msun$ out of the single metal-enriched molecular cloud. Hence, once a SPH particle exceeds the threshold density $n_{\rm H, cl}=220 \cmci$ for the formation of molecular clouds and the critical metallicity, $Z_{\rm crit}=10^{-5}\zsun$, 
%we immediately create an effective sink particle with a mass of $m_{\ast, \rm Pop~II}=3000\msun$, by accreting 
%surrounding gas particles. For the Pop~II IMF, we use a normal Salpeter IMF, $dN/d\log m\approx m^{-\alpha}$, with a slope 
%$\alpha=1.35$ over the mass range of $[1-100]\msun$. 
\par

\begin{figure*}
  \includegraphics[width=177mm]{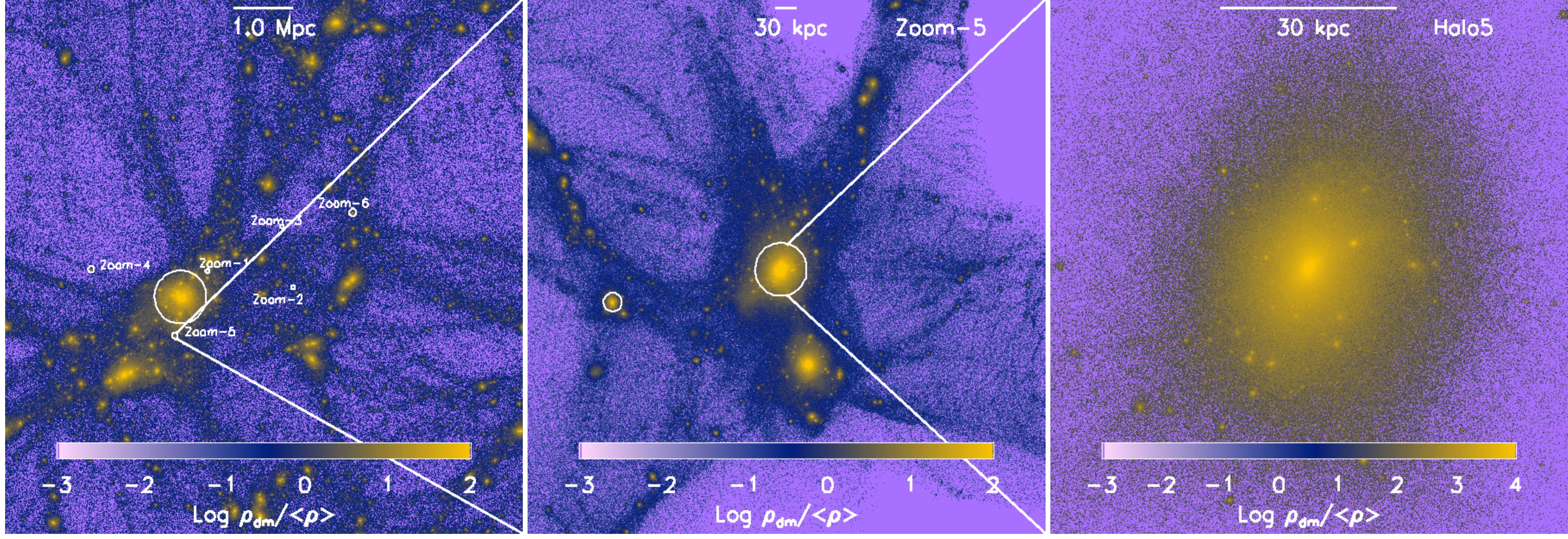}%{sedov_test.eps}
   \caption{Dark matter distribution on progressively smaller scales at $z=0$, as labeled on each panel. White 
   lines denote the region to be depicted in the next-smallest scale. {\it Left}: Final simulation output from the dark-matter only preliminary run that includes a MW like 
   halo with a mass of $M_{\rm vir}\sim2\times10^{12}\msun (z=0)$, shown as the halo surrounded by the largest white circle. Six different zoomed-in regions around the central host 
   halo are also presented as small white circles. {\it Middle:} DM distribution of the selected zoom-in region. There are two haloes in the refined region including the 
   most massive halo, {\sc Halo5}, which is zoomed in the right panel. {\it Right}: Detailed look of {\sc Halo5}. The virial radius of the halo is $\sim30$ kpc at $z=0$.}
\end{figure*}

%\subsection{Stellar Feedback}
%In this section, we describe how we implement chemical feedback and thermal feedback from SN explosions. We 
%do not include photo-heating and radiation pressure from stars for the sake of feasibility, since it would be computationally prohibitive to
%follow galaxy evolution over time down to $z=0$, while self-consistently solving the radiative transfer equation for our highly-resolved star forming regions.

\subsection{Chemical Feedback}
In this section, we describe how we implement stellar yields from SN explosions and metal diffusion into the ISM and IGM. 
We use an implementation of chemodynamics where enrichment by winds
from asymptotic giant branch (AGB) stars, and by Type~II and Type~Ia SNe are incorporated (\citealp{Wiersma2009}). At every time step, 
the mass of 11 individual elements from dying stars is computed and released into the 
neighboring medium, followed by subsequent diffusive mixing in the ISM and IGM. We have, for the first time, applied the metal 
diffusion scheme to understand the evolution of chemical abundances of dwarf galaxies in the full cosmological context.
We briefly outline the procedure here.

\subsubsection{Stellar Yields: Population~III Stars}
The low atmospheric opacities due to the absence of metals renders mass loss by stellar winds from primordial stars almost 
negligible (e.g. \citealp{Kudritzki2000}). Therefore, the final fate of Pop~III stars is determined by their initial mass. For example, massive stars in the mass 
range between $10\msun$ and $40\msun$ end their lives as conventional CCSNe, some of which are hypernovae in the range of 
$25\msun\lesssim m_{\ast}\lesssim40\msun$. Very massive stars within the particular mass range of $140\msun \lesssim m_{\ast} \lesssim 260\msun$ are expected to 
experience PISNe, powered by electron-positron pair creation. In the latter case, after central helium burning, the core temperature becomes sufficiently high
to abundantly produce electron-positron pairs, such that the star begins to collapse due to the sudden drop in radiation pressure support.
The rapid contraction in turn triggers implosive oxygen and silicon burning which produces enough energy 
to completely disrupt the star, leaving no remnant behind. If Pop~III stars are more massive than $m_{\ast}>260\msun$, they are expected to undergo 
direct collapse into a black hole without any explosion since the thermonuclear energy release is not sufficient to reverse the implosion. 
\par
Nucleosynthetic yields of CCSNe from Pop~III progenitor stars and their remnant masses are provided by \citet{Heger2010}, comprising progenitor masses of
$10-100\msun$, and exploring a range of explosion energies
between $0.3\times 10^{51}$ erg and $10^{52}$ erg. Specifically, we use yields for SN explosion energy of $1.2\times10^{51}$~erg, in the absence of 
rotation. For PISNe, we adapt the yields given in 
\citet{Heger2002}. It should be noted that 
the effects of stellar rotation and magnetic fields are not considered. E.g., some recent studies suggest that under the influence of stellar rotation 
the lowest mass resulting in a PISN can be shifted down to $75\msun$ (e.g. \citealp{Chatzopoulos2012}).    

\subsubsection{Stellar Yields: Population~II Stars}
Stars experience strong mass loss at the end of their lives, corresponding to the AGB or SN phase. The lifetime of a star is 
defined as the time when a star leaves the main-sequence and enters its red giant phase. 
Here, we employ metallicity-dependent tables of stellar lifetimes, covering values from $Z=0.0004$ to $Z=1.0$ (\citealp{Portinari1998}). 
Intermediate-mass stars ($0.8\msun\lesssim m_{\ast}<8\msun$) 
lose up to $60\%$ of their mass during the terminal AGB stage. The corresponding yields are taken from \citet{Marigo2001}, where stars as massive
as $5\msun$ are covered. Due to the low wind velocity compared to the velocity dispersion in the 
ISM, the kinetic energy input from AGB winds is neglected.
\par
Substantial amounts of metals are ejected by Type~II SNe when massive stars ($m_{\ast}\gtrsim8\msun$) explode 
at the end of their life. The yields from Type~II SNe are from \citet{Portinari1998}, in which mass loss on the main sequence 
is considered. They provide a yield set that is derived in a self-consistent manner with the AGB yields, in terms of stellar lifetimes and mass range considered. 
Due to our short hydrodynamic timesteps, $\Delta t=0.01-0.1$~Myr, compared to the timescale over which Type~II SNe 
occur, metals are released through multiple timesteps after the most massive star in a stellar cluster undergoes a SN explosion, followed 
by subsequent explosions of less massive stars. The energy associated with Type~II SNe is distributed 
onto neighboring particles as thermal energy, comprising a total of $E_{\rm SN, Pop~II}=8.6\times10^{51}$ erg. This value is obtained 
by integrating over a Chabrier IMF in the mass range of $[6,100]\msun$ for a Pop~II cluster. The SN energy 
is deposited at once in a single time step at the end of the most massive
star's life in a Pop~II cluster. A detailed description is to follow in Section~2.5.
\par
The progenitor 
mass of a Type Ia SN is thought to be between $3\msun$ and $8\msun$. Once the mass of a white dwarf remnant
exceeds the Chandrasekhar mass, either by mass transfer from a companion star or the merger of two white dwarfs, 
a Type~Ia SN is triggered. Due to the large uncertainty attributed to the details of binary evolution, determining the Type~Ia SN rate is highly complex, compared 
to the mass release by Type~II SNe and AGB stars, where mass loss simply occurs at the end of the progenitor's life. 
We employ empirical delay time functions (e.g. \citealp{Barris2006}; \citealp{Forster2006}), expressed as e-folding times, 
$\eta(t)=e^{-t/\tau_{\rm Ia}}/\tau_{\rm Ia}$ where $\tau_{\rm la}=$2 Gyr is the characteristic delay time. 
The SN~Ia rate at a given timestep $\Delta t$ is then  $N_{\rm SN~Ia}(t; t+\Delta t) = a \int_t^{t+\Delta t}{f_{\rm wd}(t')\eta(t')dt'}$, where 
$a=0.01$ is a normalization parameter and $f_{\rm wd}$ is the number of stars that have evolved into white dwarfs per unit stellar mass 
(\citealp{Mannucci2006}). The Type~Ia SN yields, based on the explosion of a Chandrasekhar-mass carbon-oxygen white dwarf, 
are taken from the spherically symmetric ``W7" model (\citealp{Thielemann2003}). The corresponding SN energy is again distributed
onto neighboring particles as thermal energy, but this feedback is likely to be less effective in disturbing the surrounding gas because 
the energy injection is distributed over billions of years.

\subsubsection{Metal Diffusion}
\par
After metals from AGB stellar winds, from Type~II and Type~Ia SNe, and from PISNe are released, they should be transported into the ISM and IGM. 
However, owing to the lack of intrinsic mass flux between SPH particles, the implementation of metal transfer in SPH simulations is nontrivial. 
Commonly, a ``particle metallicity" has been widely used, where metals are locked up into the initial neighboring gas particles, resulting in 
a very compact metal distribution around a SN explosion site. A new improved implementation has been suggested by \citet{Wiersma2009}, 
where gas metallicity is computed in terms of density $Z_{\rm sm} = \rho_{\rm Z}/ \rho$. This smoothed metallicity is suitable 
in simulating galaxy formation because gas cooling, one of the key factors in star formation, depends on gas density. Thus using
the smoothed metallicity, derived with the SPH kernel formalism, allows one to compute metallicity-dependent gas cooling in a 
more consistent way. Also, the smoothed metallicity partially accounts for the spreading of metals.
The method we adopt here, on the other hand, is a diffusion-based metallicity implemented by \citet{Greif2009}, 
where the mixing efficiency on unresolved scales is determined by the physical properties on the scale of the SPH 
smoothing kernel (\citealp{Klessen2003}).
\par
Initially, the ejected metals are distributed among the neighboring gas particles, $N_{\rm ngb}=48$, giving rise to initial particle metallicities, 
\begin{equation}
Z_i = \frac{m_{\rm metal,i}}{m_{\rm SPH}+m_{\rm metal,i}}.
\end{equation}
Then, the metals are transported by solving the diffusion equation, written in the form,
\begin{equation}
\frac{dc}{dt} = \frac{1}{\rho}\nabla \cdot ({D\nabla c}),
\end{equation}
where $c$ is the concentration of a fluid per unit mass, corresponding to the total gas metallicity in this work. We also track individual metal species 
element-by-element such as C, O, Si, Mg, Ne, Ni, and Fe. $D$ is the diffusion coefficient defined as $D=2$ $\rho$ $\tilde{v}$ $\tilde{l}$, 
where the length scale, $\tilde{l}$, is comparable to the smoothing length of the SPH kernel, $\tilde{l}=h$, and $\rho$ 
is the gas density. The velocity dispersion within the kernel, 
$\tilde{v}$, is given by
\begin{equation}
\tilde{v}_i^2 = \frac{1}{N_{\rm ngb}} \sum_j |v_i-v_j|^2.
\end{equation}
Here, $v_i$ and $v_j$ are the velocities of particles $i$ and $j$ within the kernel. 
Efficient metal mixing is achieved owing to the increased velocity dispersion of the gas as the forward shock from 
a SN explosion passes by. 
It should be mentioned that this method assumes that motions in the resolved scales cascade down to unresolved scales within which the gas 
is homogeneously mixed. The resulting velocity field is driven by a homogeneously and isotropically turbulent medium, meaning 
that three-dimensional structures within the medium are not considered.

\subsection{Thermal Feedback}

During the main-sequence stage, Pop~III and Pop~II stars emit photons that ionize and heat the surrounding medium, but 
in this work, we do not include photo-heating and radiation pressure from stars for the sake of feasibility. 
It would be computationally prohibitive to follow galaxy evolution over time down to $z=0$, while self-consistently 
solving the radiative transfer equation for our highly-resolved star forming regions. 
Once a massive Pop~III or Pop~II star dies, exploding as a Type~II SN, the explosion energy is injected into the surrounding medium, destroying 
dense gas, and thus suppressing subsequent star formation. Here, we insert SN energy as thermal energy, distributed onto
neighboring gas particles. The common problem regarding the 
thermal energy method is that gas particles heated by a SN explosion radiate their energy away too quickly, 
before making SN feedback effective. This is in part due to the unphysical 
absence of photoheating in our current simulations, which would otherwise
act to photoevacuate gas prior to the SN explosion, rendering the surrounding medium too dense. 
\par
Another problem is that
SN energy is deposited onto too much gas mass, owing to the limited resolution. In reality, one CCSN is triggered out of every $\sim100\msun$ in stars, and
the SN energy is carried by $<<100\msun$ of ejecta, while in cosmological simulations the mass of neighboring gas particles that receive the SN energy 
is much larger than the physical ejecta mass. Consequently, temperatures in 
the heated gas remain below $T\sim10^6$~K, where the cooling time is too short. Therefore, one widely used way to circumvent the 
over-cooling problem is disabling gas cooling for a few tens of Myr. In our simulations, on the other hand, we follow the strategy proposed by 
\citet{Vecchia2012}, which assigns temperatures above $10^{7.5}$~K to the heated particles, thus preventing the gas from radiatively losing its energy too quickly. 
We briefly summarize this approach here.
\par
The temperature jump of gas particles that receive SN energy is given by
\begin{eqnarray}
\Delta T &=& (\gamma-1) \frac{\mu m_{\rm H}}{k_{\rm B}} \epsilon_{\rm SN} \frac{M_{\ast}}{m_{\rm g, heat}}  \\
&=& 4.34\times10^7 \unit{K} \left( \frac{n_{\rm SN}}{1.736\times 10^{-2}\msun^{-1}} \right) \left( \frac{\mu}{0.6} \right) \nonumber \\
&\times& E_{\rm 51} \frac{M_{\ast}}{m_{\rm g, heat}}, \nonumber
\end{eqnarray}
where $\epsilon_{\rm SN}=n_{\rm SN} E_{\rm 51}\times10^{51} \unit{erg}$ is the total available SN energy per unit stellar mass and 
$E_{\rm 51}\times10^{51} \unit{erg}$ ($E_{51}=1$) is the available energy from a single CCSN event. 
The number of stars per unit stellar mass ending their lives as Type~II SNe is defined as 
$n_{\rm SN} = \int_{m_0}^{m_1}{\phi(m)} dm$ where $m_0$ and $m_1$ are the minimum 
and maximum initial mass of stars eligible for SN explosion, and $\phi(m)$ is a given IMF. 
For Pop~III clusters, the number densities of CCSNe and PISNe per stellar 
mass are $n_{\rm CCSN, Pop~III}=1.2\times10^{-2}\msun^{-1}$ ($[m_0, m_1]=[11,40]\msun$) 
and $n_{\rm PISN, Pop~III}=4.9\times10^{-4}\msun^{-1}$ ($[m_0, m_1]=[140,150]\msun$), respectively. The latter means 
that typically two PISN events happen in the $500\msun$ Pop~III cluster, $N_{\rm PISN}=n_{\rm PISN, Pop~III} M_{\ast, \rm Pop~III}\sim0.24$. 
For Pop~II clusters with the assumed Chabrier IMF, the number density of CCSNe is $n_{\rm CCSN, Pop~II}=1.73\times10^{-2}\msun^{-1}$ ($[m_0, m_1]=[8,100]\msun$). 
The total available SN energy from a single Pop~III and Pop~II cluster 
is $E_{\rm SN}= (\epsilon_{\rm CCSN, Pop~III} + \epsilon_{\rm PISN, Pop~III}) \times M_{\ast, \rm Pop~III}=2.75\times10^{52} \unit{erg}$, 
where $\epsilon_{\rm PISN, Pop~III}=n_{\rm PISN,Pop~III}\times (7\times10^{52})$ ergs,  and $\epsilon_{\rm CCSN, Pop~II} \times 
M_{\ast, \rm Pop~II}=8.5\times10^{51} \unit{erg}$, respectively. In order to assure that the heated gas reaches above $10^7$ K, making the thermal SN feedback effective in the surrounding medium, 
we release the SN energy at once when the most massive star in a cluster undergoes a SN explosion. 
\par

In the standard SPH thermal feedback implementation, SN energy is normally distributed onto neighboring SPH particles, where $N_{\rm neigh}=48$. If we heat all the neighboring particles, then the
total heated gas mass is $m_{\rm g, heat}=m_{\rm SPH}N_{\rm neigh} = 2.3\times10^4\msun$ and it renders the ratio $M_{\ast}/ m_{\rm g, heat}=0.02$ for the Pop~III and Pop~II clusters. Consequently, the neighboring gas particles would achieve a temperature jump lower than $\Delta T\sim10^{7.5}$K by an order of magnitude, making SN feedback incapable of impacting the gas. In order to increase the temperature jump, we decrease the ratio  $M_{\ast}/ m_{\rm g, heat}$ by reducing the number of neighboring heated particles to a single particle, the closest one around a stellar cluster. Heating a single gas particle ensures the ratio $M_{\ast}/ m_{\rm g, heat}$ to be of order unity.
\par

%Note that under the existence of rotation of Pop~III progenitor in a mass range of $25-40\msun$, a much stronger explosion
%may be expected, likely resulting in a hypernova (e.g. \citealp{Fryer2000}; \citealp{Nomoto2006}). However, due to
%the uncertainty in the degree of spin of Pop~III stars, which is subject to stellar winds, gravitational and
%hydromagnetic instabilities (e.g. \citealp{Stacy2011}), we only consider conventional CCSNe by fixing the explosion
%energy and metal yield.
Additionally, we use a timestep-limiter such that the ratio of timesteps of neighboring SPH particles cannot be larger than a given factor,
here adopting a value of 4 (\citealp{Saitoh2009}; \citealp{Durier2012}). The implementation is essential for the correct treatment of SN explosions,
especially for high resolution multiphase simulations with individual time-steps, where the hot gas ($T>10^{7.5}$~K) is
located near cold, dense gas ($T<10^4$~K), resulting in a large difference in their timesteps that eventually could lead to a large integration error.
Also, we consider a timestep-update where all neighboring particles around a SN explosion site
become active particles at the time of energy injection (\citealp{Vecchia2012}), allowing them to
immediately react to a sudden SN event.

\begin{figure*}
\centering
  \includegraphics[width=180mm]{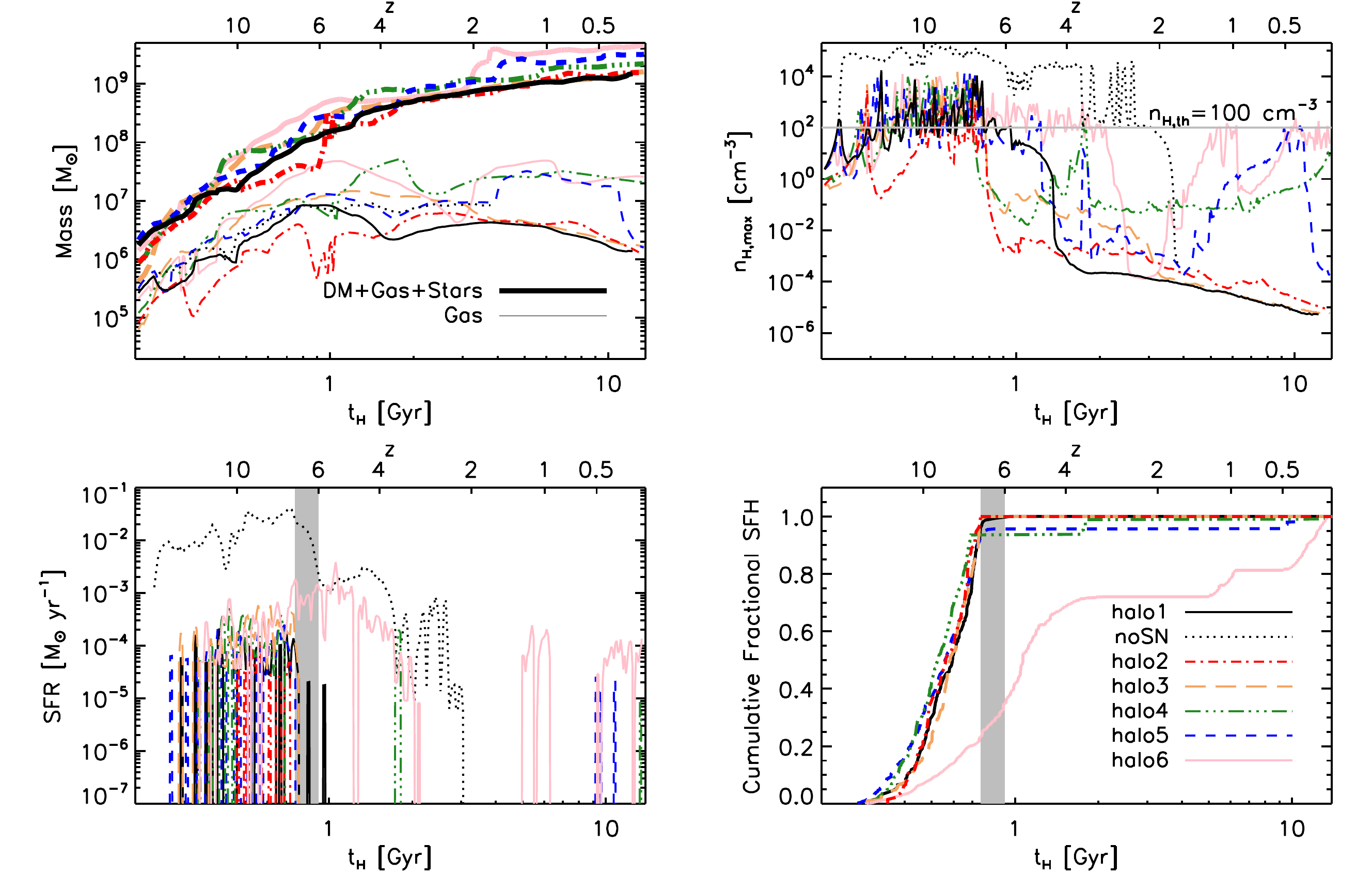}%{sedov_test.eps}
   \caption{Time evolution of the UFD analogs. {\it Clockwise from upper left}: total and gas mass, 
   the maximum hydrogen density, cumulative fractional star formation history, and star formation rates within the virial radius of the haloes. The UV background 
   is introduced at $z=7$ starting from zero strength and gradually increasing to the full amplitude at $z=6$, depicted as the grey shaded region in the bottom 
   plots. Different colors and line types indicate the evolution of each UFD analog. The dotted line indicates results when no SN feedback is included in the
case of {\sc Halo1}. Relatively less massive haloes ({\sc Halo1, Halo2}, and {\sc Halo3}), whose halo mass is less than $2\times10^9\msun (z=0)$, form {\it all} stars 
   prior to reionization. On the other hand, in {\sc Halo4} and {\sc Halo5}, late starbursts are triggered mainly driven by mergers with other haloes. The most 
   massive halo, {\sc Halo6}, forms only about $30\%$ of stars before reionization and star formation is continuously triggered under the influence of the UV background. 
   Star formation continues well past $z=6$, illustrating that reionization and the impact of SNe are insufficient to suppress SF in {\sc Halo 6}. {\sc Halo 6} is thus 
   our prime candidate for an analog of a gas-rich, low-mass dwarf system, such as, e.g., Leo P.}
\end{figure*}
  
  \begin{figure}
\centering
  \includegraphics[width=85mm]{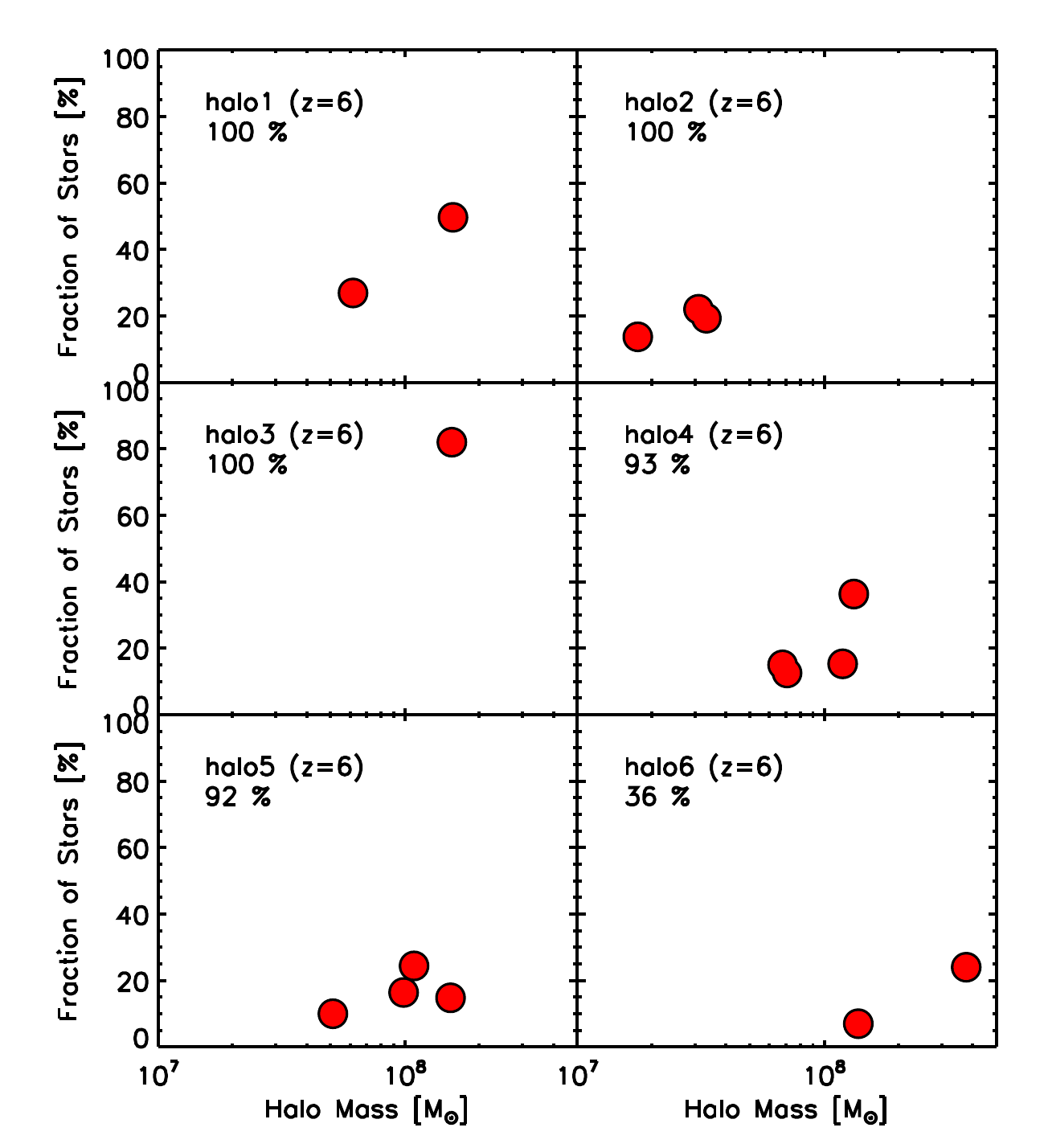}%{sedov_test.eps}
   \caption{Fraction of stars within progenitor haloes found at $z=6$ that contribute more than 5\% of the total stars within the virial radius of the simulated galaxies 
  at $z=0$. The percentage indicates the fraction of stars formed at this time relative to the final stellar mass at z=0.
  Prior to $z=6$, in all cases stars were formed in multiple different progenitors, not in a single primary halo. This means that a 
substantial amount of stars have originated from the hierarchical accretion of different haloes, rather than being formed in-situ. The ability of the simulations resolving small halos ($M_{\rm vir}\lesssim10^8\msun$) at $z>6$ allows us to track hierarchical structure formation at high-$z$. At early epochs ($z\gtrsim6$), star formation
occurs simultaneously in multiple small halos ($M_{\rm vir}\lesssim10^8\msun$, at $z>6$), merging into a primary halo at later times. 
In all cases, except for the most massive {\sc Halo6}, the haloes almost stop forming stars at $z=6$. Prior to this point multiple star-forming small haloes are present. Consequently, 
star formation of the simulated UFDs is found to be a combination of in-situ star formation in a primary halo and significant 
contribution from small progenitor haloes that eventually merge into a primary halo. Note, however, that {\sc Halo6} continues to form stars until $z=0$, meaning
81\% of its stars are formed in-situ within virial radius of a single primary halo. This points to a notable 
distinction in the stellar mass build-up of UFDs vs. low-mass dwarfs.}
\end{figure}

\section{Simulation Results}
In the following, we present our simulation results concerning the chemical properties of low mass reionization 
relics, in comparison to the observed properties of the lowest mass galaxies about the MW. 
In Section~3.1, we consider the star formation history of the simulated 
galaxies over cosmic time down to $z=0$. 
% GB modified - to indicate straight away that we are identifying Halo 6 as a different thing 
In Section~3.2, we focus on {\sc Halo 6}, the most massive simulated halo, 
in order to discuss residual gas at $z=0$ and analogs of low-mass, gas-rich 
dwarfs, such as Leo P, Leo A, Leo T and DDO210. 
Next, we explain the general trend of metal yields from Pop~III and Pop~II stars 
in Section~3.3, and the resulting chemical abundances of the UFDs in Section~3.4. We further discuss how to provide 
insight into the SFHs and accretion timescales of the observed LG UFDs by comparing with our simulated UFD analogs. 
Finally, in Section~3.6, we present the global galaxy properties at $z=0$ and compare our results with other theoretical 
studies and observations. All the simulated dwarfs reside at distances from 0.6$-$3.7 Mpc from the central galaxy at 
$z=0$ (see, Figure~1). As such, environmental factors (tides, ram pressure stripping, UV from central galaxy)
 are not relevant to the evolution of these galaxies.

\begin{figure*}
  \includegraphics[width=177mm]{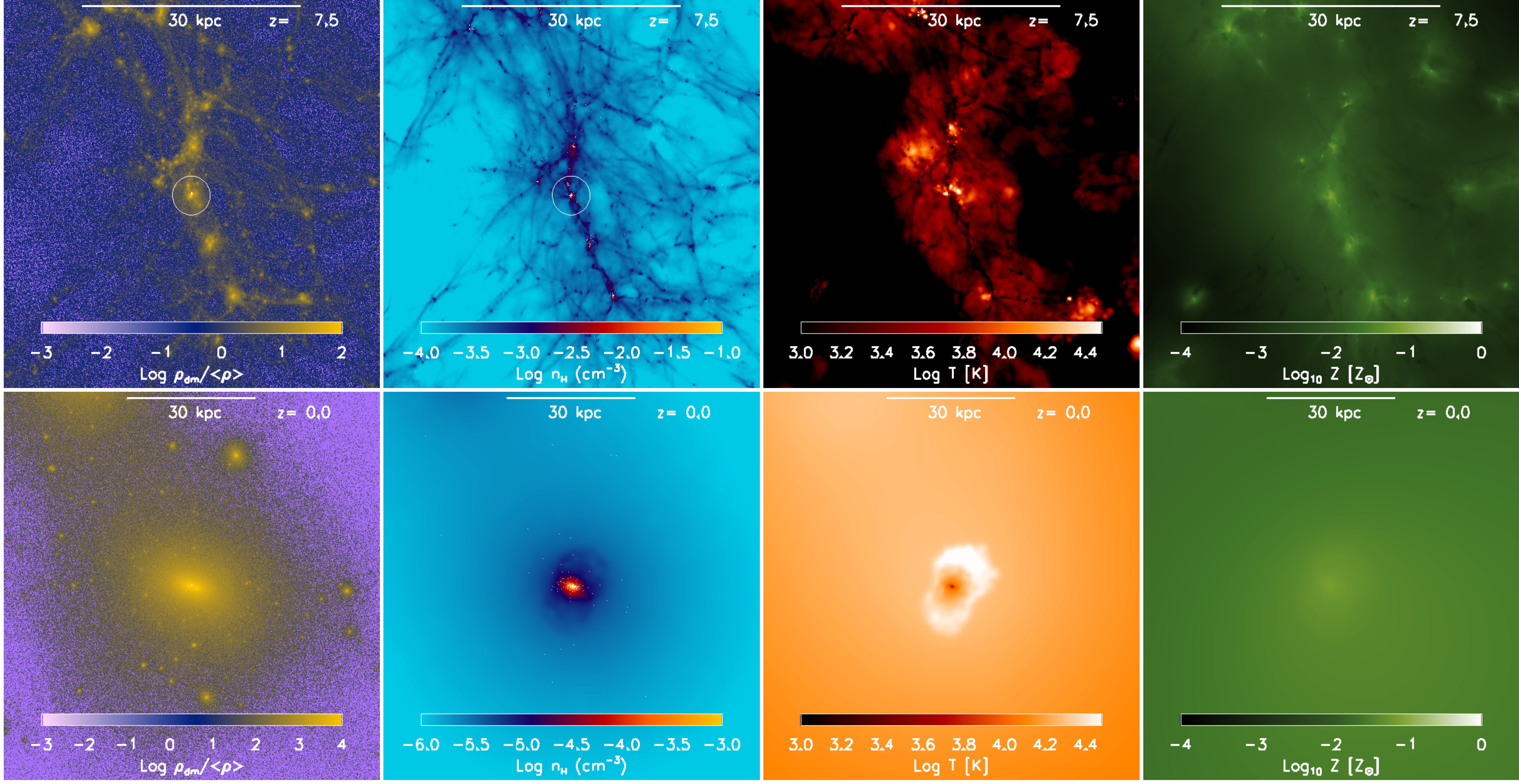}%{sedov_test.eps}
   \caption{Visualization of the zoom-in region, centered on {\sc Halo4}, at $z=7.5$ ({\it top panels}) and at $z=0$ ({\it bottom panels}), respectively. 
   {\it From left to right}: each panel displays dark matter over-density relative to average density in Universe, hydrogen number density, gas temperature, and gas metallicity. The 
   radius of the primary halo, the most massive progenitor at a given time, is denoted as the white circle in the dark matter over-density map 
   (upper left-most panel). At $z=0$, {\sc Halo4} still forms stars, some of which explode as SNe. The gas heated by the SN explosions is 
   clearly shown in the temperature map at $z=0$. We see that, at $z=7.5$, stars are formed and explode as SNe within multiple progenitor haloes and that are then assembled into the primary halo. As a result, the galaxy at $z=0$ is composed of a significant fraction stars that have originated 
   in different haloes that were later cannibalized by the primary.}
\end{figure*}

\subsection{Star Formation History}
\subsubsection{SFH of the Primary Dwarf Halo (i.e. the most massive halo in its merger history)}
Figure~2 shows the time evolution of the zoomed-in low mass haloes in each refined region 
({\it clockwise from upper left-hand}): the total and gas mass, the maximum hydrogen 
number density, cumulative fractional SFH, and star formation rate (SFR). These quantities are computed based on the properties of particles within the 
virial volume of the haloes. 
Most of the haloes begin to grow from minihaloes with a mass of $M_{\rm vir}=10^5-10^6\msun$ at $z\gtrsim15$, inside of which
early Pop~III stars form. The emergence of first star formation differs from halo to halo: the lower the halo mass, 
the later Pop~III stars begin to form. This is because a halo must reach a minimum mass required 
to host gas with a molecular hydrogen fraction of $f_{\rm H_2}\gtrsim10^{-4}$, above which $\rm H_2$ 
cooling is efficient (e.g. \citealp{Tegmark1997}; \citealp{Bromm2002}). We see that Pop~III stars emerge at $z\lesssim11$ 
in the less massive haloes ({\sc Halo1, Halo2}), while the more massive haloes 
({\sc Halo4, Halo5, Halo6}) commence Pop~III star formation before $z=15$. 
The final virial masses at $z=0$ are less than $M_{\rm vir}\sim1.6\times10^9\msun$ for 
three haloes {\sc Halo1, Halo2, Halo3}, while the others are slightly more massive with
$M_{\rm vir}\sim2.1\times10^9$ ({\sc Halo4}), $M_{\rm vir}\sim3.1\times10^9\msun$ ({\sc Halo5}), 
and $M_{\rm vir}\sim4\times10^9\msun$ ({\sc Halo6}), respectively.

\par
Note that early on, prior to reionization, star formation proceeds in an episodic fashion. This is a consequence of regulated star formation: 
the gas within the shallow potential well of a dwarf can easily be disrupted and evacuated by SN feedback, and it takes 
time for the halo gas to be replenished. This time delay gives rise to the episodic, bursty star formation with timescales of the order of a few $\sim10$ Myr. 
As explained in Section~2, we gradually increase the external UV background, starting from $z=7$ to its full strength by $z=6$, indicated by the grey 
shaded vertical region in Figure~2; afterwards, the background is present at full amplitude down to $z=0$. 
\par

The star formation history of the simulated galaxies strongly depends on their halo mass. 
Quenching of star formation by reionization is clearly reflected in the noticeable drops in SFR during $z\sim7-6$ in the less massive haloes, {\sc Halo1, Halo2}, and {\sc Halo3}. 
Among them, {\sc Halo1} continues to form stars for about $\sim$30 Myr beyond $z=7$, while {\sc Halo2} and {\sc Halo3} exhibit
sharply truncated star formation as soon as the UV background is introduced. This residual star formation in {\sc Halo1} is due to 
the existence of dense gas, self-shielded from the UV background.
Although two other haloes, {\sc Halo4} and {\sc Halo5}, also experience SFR quenching by the UV background, they exhibit a rebirth
of star formation at late epochs, below $z=4$.  Gas inflow in {\sc Halo5} is found to be driven by mergers with other haloes
at $z\sim1.8$, such that the primary progenitor of {\sc Halo5} grows in mass by a factor of 5, leading to the ignition of star formation at 
$z\lesssim0.5$. {\sc Halo4} shows late starbursts at $z\sim3.9$ and $z\sim0$, as well. 
\par
Interestingly, the most massive halo 
({\sc Halo6}) overcomes the effect of negative feedback from UV photoheating, since the halo already reaches a mass of 
$M_{\rm vir}=2.5\times10^8\msun$ at $z=7$, which is massive enough not to significantly be affected by external radiation. 
Therefore, stars can continuously form down to $z\sim3$, followed by bursty star formation until $z=0$. 
These late time bursts of star formation could be unlikely to occur if the simulated galaxy, {\sc Halo6}, was accreted by the 
MW-like host at early times. As such, star formation would be suppressed by environmental factors. This implies that 
low-mass dwarfs ($M_{\rm vir}\gtrsim2\times10^9\msun$ at $z=0$) in the 
outskirts of the MW are likely to show signatures of late time bursts of star formation. Or, such ongoing star formation 
and high stellar content ($M_{\ast}\sim10^6\msun$ at $z=0$) indicate that it is more reasonable to consider 
{\sc Halo6} as low-mass dwarf analogs such as Leo~P or Leo~T in the field rather than local UFDs.  This is discussed 
further in the next section.

%Furthermore, if such 
%massive UFDs are located within the virial radius of the MW, it is expected that, to explain their uniformly 
%quenched SFH, they must have been capture by the MW at early times.

\par
To explore the relative importance of SN feedback, compared to reionization, for the suppression of star formation, we perform a comparison 
simulation {\sc noSN} (see dotted lines in Figure~2). This run is identical to {\sc Halo1}, except that the thermal SN feedback is excluded, but metals are ejected 
from stars to model the transition of star formation from Pop~III to Pop~II. We find that, while reionization plays an important role in 
inhibiting star formation, SN feedback is crucial for the ultimate quenching of UFDs. Although the UV background 
is turned on in the same way as before ({\sc Halo1}), no quenching of star formation is found this time, implying that reionization alone is insufficient 
to completely suppress star formation in the absence of SN feedback. We should note that we terminate the run {\sc noSN} at $z\sim2$ for the sake of 
computational economy.
\par

The bottom right-hand panel of Figure~2 illustrates that all haloes form more than $90\%$ of stars prior to reionization, with the noticeable exception of {\sc Halo6}. 
Following \citet{Ricotti2005}, our five galaxies can be considered as ``true fossils", as more than $70\%$ of their stars formed before reionization. 
On the other hand, {\sc Halo6} only forms $\sim30\%$ of stars before reionization and continues to form 
stars down to $z=0$, and hence it could be categorized as a ``polluted fossil".
\par
\subsubsection{SFH of the Smaller Progenitor Haloes}
Near the end of reionization, $z\sim6$, the simulated haloes, except {\sc Halo6}, almost stop forming stars such that 
{\sc Halo1, Halo2,} and {\sc Halo3} complete 100\% of their star formation. Also, {\sc Halo4} and {\sc Halo5} already contain 
more than 92\% stars by $z\sim6$. Interestingly, we find that simulated analogs have experienced 
multiple mergers between haloes, meaning that a substantial fraction of stars were formed in small progenitor 
haloes ($M_{\rm vir}\lesssim10^8\msun$) at high redshift ($z\gtrsim6$) and assembled into a primary halo at later times. 
Figure~3 shows a fraction of stars in progenitor haloes at $z=6$ that contain more than 5\% of total stars found within 
a virial radius of the simulated analogs at $z=0$. The percentage indicates the fraction of stars formed at this time 
relative to the final stellar mass at $z=0$.
\par 
For instance, there were two massive progenitor haloes of {\sc Halo1} at $z=6$, $M_{\rm halo}=1.6\times10^8\msun$ 
and $M_{\rm halo}=6\times10^7\msun$, containing 50\% and 27\% of stars.
As the number of progenitors that contribute non-negligible amount of stars increases, the fraction of stars 
formed in a primary halo, the most massive halo, declines as seen in {\sc Halo2}, {\sc Halo4}, and {\sc Halo5} 
of Figure~3. This means that the stellar mass growth of the simulated UFD analogs is a combination of both in-situ 
star formation in a primary halo and stellar accretion via mergers of smaller progenitor haloes.
\par
The ability of the simulations resolving small haloes ($M_{\rm vir}\lesssim10^8\msun$ at $z\gtrsim6$), 
allows us to track down hierarchical structure formation to high redshifts, making it difficult to specify a ``primary" halo, 
particularly for the relatively less massive analogs ($M_{\rm vir}<2\times 10^9\msun$), in which progenitor haloes were 
formed in low-density peaks. This confusion lessens for the relatively massive halo, i.e. {\sc Halo6}  ($M_{\rm vir}\gtrsim 4 \times 10^{9}\msun$ at $z=0$)
  as their progenitor haloes grow rapidly by mergers and accretion since they were originated from high-density peaks. As such, it is much 
  easier for relatively massive haloes to identify a primary halo. 
  This is clearly seen in {\sc Halo6}, where stars actively form in two progenitor haloes before $z=6$, but  
afterwards stars form in a single halo, leading to 81\% of in-situ stars in {\sc Halo6}. 
On the other hand, unlike {\sc Halo6}, the other five simulated haloes almost complete star formation by $z\sim6$ - before stars were formed 
in multiple small progenitor haloes.  The fraction of stars formed in-situ (within a primary halo) varies from halo to halo:  
37\%, 17\%, 61\%, 31\%, and 17\% from {\sc Halo1} to {\sc Halo5}. This suggests that there is a notable distinction in 
the stellar mass build-up of UFDs vs. low-mass dwarfs.

\begin{deluxetable}{c c c}
%\centering
\tablecolumns{4}
\tablewidth{0.4\textwidth}
\tablecaption{Summary of gas-rich nearby low-mass dwarfs. }
\tablehead{\colhead{Dwarfs} & \colhead{$M_{\ast}[10^5\msun]$} & \colhead{$M_{\rm HI}[10^5\msun]$}}
\startdata
\vspace{0.07cm}
Leo A & 60 & 110 \\
\vspace{0.07cm}
DDO 210 & 16 & 41  \\
\vspace{0.07cm}
Leo P &  5.6&  9.3\\
\vspace{0.07cm}
Leo T & 1.4 & 2.8\\
\vspace{0.07cm}
{{\sc Halo6}} & 8.8 & 44 (warm)/0.59(cold)
%\vspace{0.07cm}
%{{\sc Halo6} (z=3)} & 5.9 & 20& &
%\vspace{0.07cm}
%\vspace{0.07cm}
%{\sc Halo6} &8.8& 49 & &
\enddata
\vspace{-0.1cm}
\tablecomments{Column (1): the names of the observed gas-rich low-mass dwarfs and {\sc Halo6}. 
Column (2): stellar mass in $10^5\msun$. 
Column (3): neutral hydrogen mass in $10^5\msun$. The data is from \citet{McConnachie2012} for 
Leo~A, DDO 210, Leo~T and \citet{McQuinn2015} for Leo~P.}
\end{deluxetable}

\subsection{Residual Gas: Halo 6 as an analog of a low-mass, gas-rich dwarf}
% make an intro for Leo P etc. 
Here we discuss whether any of our simulated halos are reasonable analogs to recently discovered low-mass, gas-rich dwarfs, such 
as Leo P, Leo T, Leo A and DDO 210. We focus specifically on {\sc Halo6}, which, being isolated and slightly more massive than the other 
dwarfs in our simulation suite, continues to form stars until z=0.

Whether or not a halo can form stars at a given time within the simulation can be illustrated by the maximum gas density achieved inside the 
virial radius of the simulated haloes, shown in the upper right-hand panel of Figure~2. In this panel, we compare the maximum hydrogen 
number density with the adopted density threshold of $n_{\rm H}=$100 $\rm cm^{-3}$ (solid horizontal line), above which gas can form stars. 
The maximum gas density can get as high as $n_{\rm H}\sim10^4 \rm cm^{-3}$ at high redshift $z\gtrsim7$. 
The final maximum gas density at $z=0$ is heavily dependent on the specific SFH experienced by a galaxy. 
For example, a majority of gas is evacuated both by the UV background and SN feedback in the low mass haloes ({\sc Halo1, Halo2, Halo3});
hence, little gas has remained, eventually resulting in the very low maximum density of $n_{\rm H}\sim10^{-5} \rm cm^{-3}$ at $z=0$. Even though {\sc Halo5} 
is massive enough to form stars at late epochs, SN feedback from late starbursts significantly evacuates the gas from the halo, such that the maximum density is 
as low as $n_{\rm H}\sim10^{-4} \rm cm^{-3}$ at $z=0$. Meanwhile, we identify high density gas with $n_{\rm H}\gtrsim$ 10 $\rm cm^{-3}$ in {\sc Halo6} 
even at the end of the simulation.

The total residual gas mass of low mass UFDs, both hot and cold gas, is a few $\sim10^6\msun (z=0)$ (see Table 1), but for 
{\sc Halo6} it can be as high as $2.6\times10^7\msun (z=0)$. We find that the ionized gas fraction at $z=0$, defined as gas with a free electron fraction of $f_{\rm elec}\gtrsim0.99$, is 
$\sim54\%$ in {\sc Halo6} and $\sim74\%$ in {\sc Halo4}, whereas the gas is completely ionized in the other haloes. 
Figure~4 illustrates the morphology of the refined region, centered on {\sc Halo4}, at $z=7.5$ (top panels) and at $z=0$ (bottom panels), respectively. From left to right, each panel shows dark matter over-density, hydrogen number 
density, gas temperature, and gas metallicity. At $z=7.5$, given the distribution of the heated and polluted gas 
by the SN explosions, stars are formed not only in the main halo, denoted as the white circle at the center of the top panels, but also out of gas in multiple progenitor haloes that eventually have been assembled into the main halo at $z=0$. Owing to the ongoing star formation at $z=0$, the gas in {\sc Halo4} is significantly heated up to T $\gtrsim10^5$ K, as shown in the temperature map.
\par

Interestingly, we find that a substantial amount of the residual gas of {\sc Halo6} at $z=0$ exists in a neutral phase. The mass of 
warm neutral gas, defined as $T<5000$ K and $n_{\rm H}>0.5$ $\rm cm^{-3}$, is $M_{\rm HI}\sim4.6\times10^6\msun (z=0)$, corresponding to 20\% of the total gas. 
Also, there is a cold gas reservoir with $5.9 \times 10^4\msun$ ($T<1000$ K and $n_{\rm H}>10$ $\rm cm^{-3}$), which serves as the gas supply for star formation.
To date, neutral gas has not been detected in local UFDs ($\rm M_\ast < 10^5 \Msun$, $z=0$), indicating that {\sc Halo6} is likely more representative of slightly more massive 
systems. In particular, the properties of {\sc Halo6} at z=0 are similar to those of gas-rich low-mass dwarfs like 
%The absence of observations of neutral gas in local UFDs within the virial radius of their hosts 
%conflicts with our finding of the neutral gas reservoirs in {\sc Halo6}. This means that the properties of {\sc Halo6} indicate that instead of UFDs it is likely an analog of low-mass dwarfs, such as 
 Leo~T ($M_{\rm HI}\sim2.8\times10^5\msun$), Leo~P ($M_{\rm HI}\sim8\times10^5\msun$), DDO~210 ($M_{\rm HI}\sim4\times10^6\msun$), 
 or Leo~A ($M_{\rm HI}\sim1.1\times10^7\msun$)  (see, Table~2; \citealp{McConnachie2012}; \citealp{McQuinn2015})
 %which are found to be gas rich (e.g.,
  %\citelap{McConnachie2012}; \citealp{Geha2012}; \citealp{Bradford2015}; \citealp{Stierwalt2015}; \citealp{McQuinn2015}).
  \par
  This work could suggest that the observed gas-rich, low-mass dwarfs listed in Table~2 are explainable within a cosmological context if they have not been subjected 
to environmental effects (ram pressure stripping, etc) for the bulk of their evolution. Therefore, the existence of dense gas, not seen in the UFDs within the virial 
radius of the MW, could be indirect evidence of the role of the environmental effects on quenching star formation (e.g. \citealp{Wetzel2015}). 
We find that {\sc Halo6} already forms $\sim$30\% of stars at $z=6$, corresponding to stellar mass of $M_{\ast}=2.6\times10^5\msun$ ($z=6$). It indicates that 
even if {\sc Halo6} was quenched after reionization, it would not look like a UFD ($M_{\ast}<10^5\msun$ at $z=0$), meaning that there likely is a hard upper 
limit on the UFD halo mass.
   
%These results indicate that the existence of dense gas, not seen in the UFDs within the virial radius of the MW, could be 
%indirect evidence of the role of the environmental effects on quenching star formation (e.g. \citealp{Wetzel2015}). 

\subsection{Metal Enrichment}
%\subsubsection{Self-enrichment vs. external enrichment}
%VB-comment: since there is only one subsection here, we can as well omit it; makes only
%            sense, if we have multiple subsections
Given that the simulated galaxies have experienced multiple mergers with other haloes over cosmic time, it is expected that some
fraction of stars could have formed in different haloes and been accreted later onto the primary halo, i.e.
the most massive progenitor at any given time. In order to distinguish the properties of stars formed in-situ in a 
primary halo, from those of stars formed externally and accreted at later epochs, in Figure~5 we compare the stellar 
metallicity as a function of formation time. In the top panels, we plot all stars inside the virial radius of {\sc Halo1} ({\it left}) 
and {\sc Halo6} ({\it right}) at $z=0$, 
regardless of their formation site. In the lower panels we compare with stars formed only within the primary halo (in-situ). Note that each 
symbol (reversed triangle) indicates a single Pop~II stellar cluster. 
\par

We note that about $40\%$ and $19\%$ of stars in {\sc Halo1} and {\sc Halo6} at $z=0$ have originated from different progenitor haloes and were assembled 
into the primary halo at later times via mergers. We can further classify these externally formed stars into two categories: (1) stars formed out of gas 
that is externally metal-polluted due to the proximity of neighboring haloes hosting SNe; and (2) stars born within a halo that 
is independently massive enough to form Pop~III stars and subsequent generations of stars (in-situ). The latter case is similar to 
a primary halo, which contains Pop~III stars and subsequently formed stars via self-enrichment. 
A striking difference evident in Figure~5 is how broadly stellar metallicity is distributed for all stars, whereas it is concentrated 
between $\rm [Fe/H]=-2$ and $\rm [Fe/H]=-3$ for the in-situ sample, except one stellar cluster 
with $\rm [Fe/H]=-3.7$. The absence of low metallicity stars ($\rm [Fe/H]\lesssim -3$) among the in-situ sample indicates
that low metallicity stars have preferentially come from different progenitor haloes, and later acquired by the main halo through mergers.
\par
This can be explained as follows. When a Pop~III star triggers a SN explosion, the surrounding gas is heated and evacuated 
from the host halo, locally suppressing further star formation for a certain time period. Detailed studies regarding the duration of this temporary lull in
star formation in the wake of a Pop~III SN have suggested that it is of the order of a few 
10~Myr for a conventional SN energy of $10^{51}$ ergs. Or, it lasts up to a few hundred Myr for a powerful 
PISN explosion with $10^{52}$ ergs, within a host halo of $M_{\rm vir}\sim10^5-10^6\msun (z\gtrsim10)$ 
(e.g. \citealp{Greif2009}; \citealp{Ritter2012}; \citealp{Jeon2014}).
\par 
During this period of inhibited star formation, metals are propagated 
into the ISM and IGM, and can be mixed with pristine gas residing in neighboring minihaloes. If the gas density in 
such polluted neighboring haloes happens to be high enough to form stars, metal-enriched Pop~II stars would be born. 
Because only small amounts of metals would be delivered to neighboring haloes, the metallicity of the gas that is located in
the central region of the externally enriched halo would be low, giving rise to the formation of low metallicity stars. Note that owing to the compactness 
of the early Universe, the typical distance between minihaloes with $M_{\rm vir}\sim10^5\msun (z\gtrsim10)$ is
$\sim300-500$~pc, making it possible to transfer metals to neighboring haloes (\citealp{Smith2015}).
\par

\begin{figure}
\centering
  \includegraphics[width=88mm]{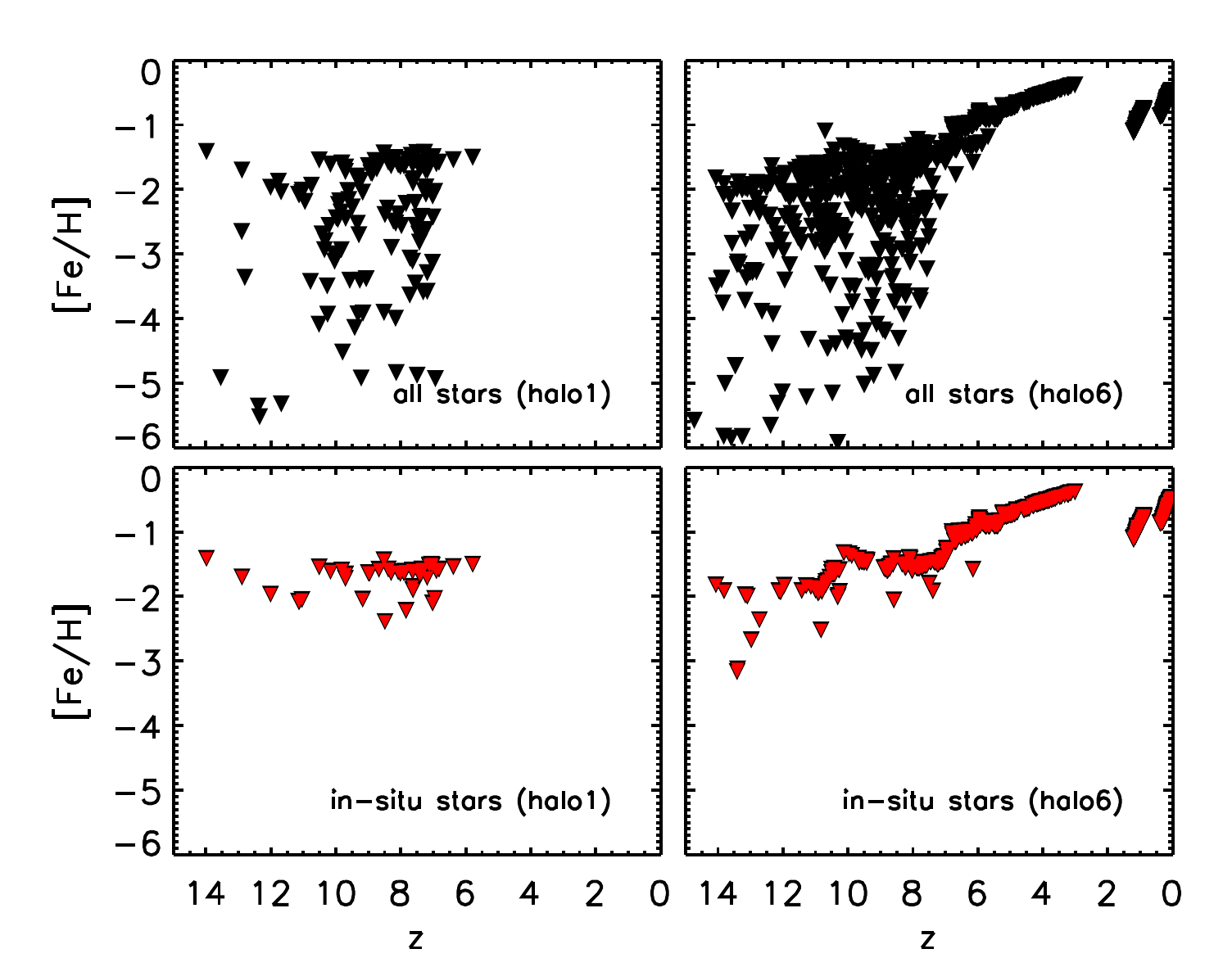}%{sedov_test.eps}
   \caption{Externally vs. internally formed stars. {\it Top panel:} stellar
   metallicity vs. formation time for {\it all} stars located within the virial radius of a halo at $z=0$.
   {\it Bottom panel:} subset of stars formed {\it in-situ} in the primary progenitor of a halo.
   The left and right panels are for {\sc Halo1} and {\sc Halo6}, respectively.
   %The vertical lines on the top axis mark the formation time of Pop~III stars.
   The absence of low metallicity stars
   ($\rm [Fe/H]\lesssim-3$) among the in-situ sample indicates that low metallicity stars have originated from
   different progenitor haloes and were were assembled into the primary halo at later epochs via mergers. Specifically, such low metallicity stars are polluted
   via external enrichment, where metals originate in neighboring haloes that host Pop~III SNe. These low metallicity stars only form at early times $z\gtrsim6$ 
   when the cross-pollution is efficient due to the proximity between haloes. This finding is a general feature in all of the simulated analogs. The same plots for 
   the other four haloes are presented in Appendix.}
\end{figure}

\begin{figure}
\centering
  \includegraphics[width=85mm]{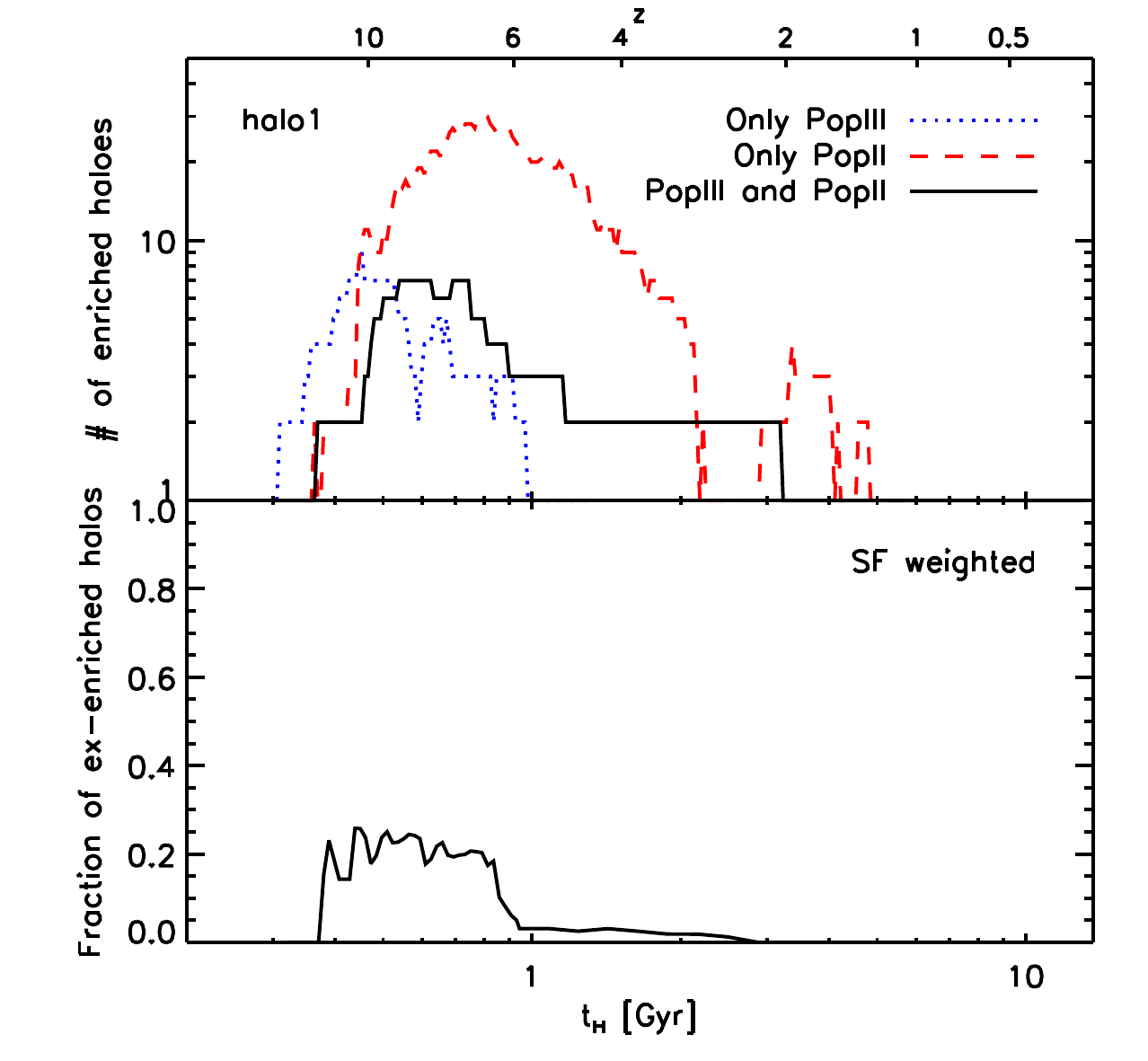}%{sedov_test.eps}
   \caption{{\it Top panel:} Number of haloes that are eventually incorporated into {\sc Halo1}, divided according to their metal-enrichment path. The blue-dotted
   line shows the number of haloes that only host Pop~III stars or their remnants, while the red-dashed line depicts haloes with only Pop~II stars. Finally, the
   black solid line indicates haloes that host both Pop~III and Pop~II star formation.
   The substantial number of Pop~II-only haloes (see red dashed line) implies that a large fraction of haloes are externally polluted with metals
   from neighboring haloes that host SNe.
   {\it Bottom panel:} The fraction of externally enriched halos weighted by star formation in each halos. As stated above, even though there are 
   a number of externally polluted halos, they tend to host only one or multiple star clusters, making a small percentage contribution to the total star formation. 
   Therefore, effectively $\sim$20\% of star-forming halos at $\gtrsim6$ were polluted via external metal-enrichment. These trends are shown in all of the simulated analogs.}
\end{figure}

The bottom-left panel of Figure~5, on the other hand, implies that the in-situ stars are likely born in an environment that 
is internally polluted, thus experiencing self-enrichment. The primary progenitor halo forms the first Pop~III stars at 
$z\sim14$ and the following Pop~III SN explosion 
leads to a suppression of star formation for $\sim50$~Myr. Afterwards, 
the ejected gas is replenished along the cosmic web, and the central gas again reaches 
densities high enough to form Pop~II stars. We find that in a given halo, only one or two Pop~II clusters 
are formed out of gas directly polluted with metals from Pop~III SNe, implying that only a few Pop~II stars will preserve 
the intrinsic chemical signatures of Pop~III stars (\citealp{Ji2015}).
While {\sc Halo1} stops forming stars at $z\sim5.5$, {\sc Halo6} continues to form stars down to $z=0$, steadily 
self-enriching the gas within the primary progenitor halo. As shown in the top panels of Figure~5, low metallicity stars 
($\rm [Fe/H]\lesssim10^{-3}\zsun$) tend to be generated at early times before $z=6$ when the cross metal-pollution is efficient 
due to the proximity between low mass halos with $M_{\rm vir}<10^8\msun$. 
\par

In order to estimate how many externally enriched progenitors contribute to the formation of the simulated dwarf galaxy at $z=0$, we present in Figure~6 the number of haloes
with different metal-enrichment paths, that end up in the primary halo ({\sc Halo1}) through mergers. The red-dashed line indicates the number of haloes that host 
only Pop~II stars without the presence of any Pop~III stars or their remnants, implying that they have been externally polluted. The 
blue-dotted and black-solid lines correspond to haloes hosting Pop~III only, or both Pop~III and Pop~II stars, respectively. 
The displacement between the blue-dotted and black-solid lines can be understood as the time delay when Pop~III star formation transitions to Pop~II, in haloes eventually
hosting Pop~III remnants and Pop~II stars. This time delay turns out to be about a few tens of Myr. 
\par
Interestingly, the number of 
externally enriched haloes increases up to $\sim30$ until $z\sim6$, as the ISM and IGM are enriched with metals from Pop~III and Pop~II SNe, but their number 
declines after reionization, mainly due to the inability of forming stars in low-mass haloes from gas heated by reionization. We point out that as global metal enrichment progresses, no Pop~III-only haloes exists $\sim$1~Gyr after the big bang. Even though there are a number of externally-polluted haloes prior to $z=6$, they only 
host a single or multiple star clusters, providing the small contribution to overall star formation. In order to estimate the effective fraction of externally 
polluted haloes, in the bottom panel of Figure~6, we present the fraction of externally enriched halos weighted by the number of stars within each 
halos. About $\sim20\%$ of haloes experience the cross metal-pollution between $z\sim11$ and 
$z\sim6$ in {\sc Halo1}. Note that all these findings are general features in all of the simulated analogs. 

\begin{figure}
\centering
  \includegraphics[width=90mm]{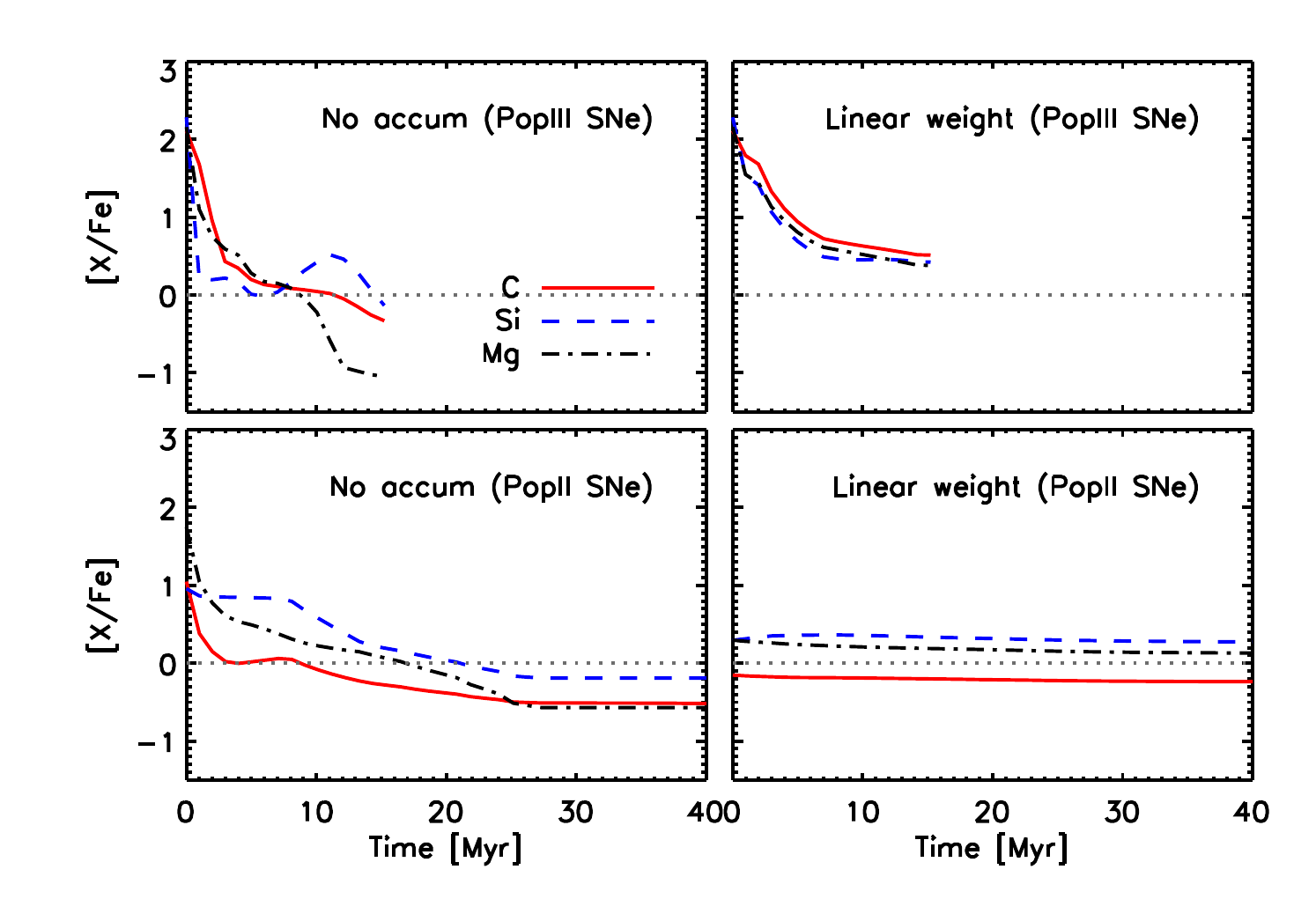}%{sedov_test.eps}
   \caption{Evolution of metal yields of Type~II SNe from Pop~III ({\it top panels}) and Pop~II stars ({\it bottom panels}).
   We assume two cases: (1) metals from individual SN events immediately propagate onwards, and thus do not accumulate in
   the high density region at the center of a halo ({\it left}); or (2) their transport is slow, such that metals pile up in the central region, 
  giving rise to an effective yield ({\it right}). For the former case, subsequently forming stars are sampling
   the metal yields from the most recent SN only, whereas the latter case means that next-generation stars will be born out of 
   gas polluted with metals from multiple SNe. Note that Pop~III yields are displayed only up to $\sim$ 15 Myr, corresponding to the
   lifetime of a $10\msun$ Pop~III star, the lowest mass considered in this work. The Pop~II yields extend to longer times, in accordance with their
lower masses.}
\end{figure}

\begin{figure*}
  \includegraphics[width=120mm]{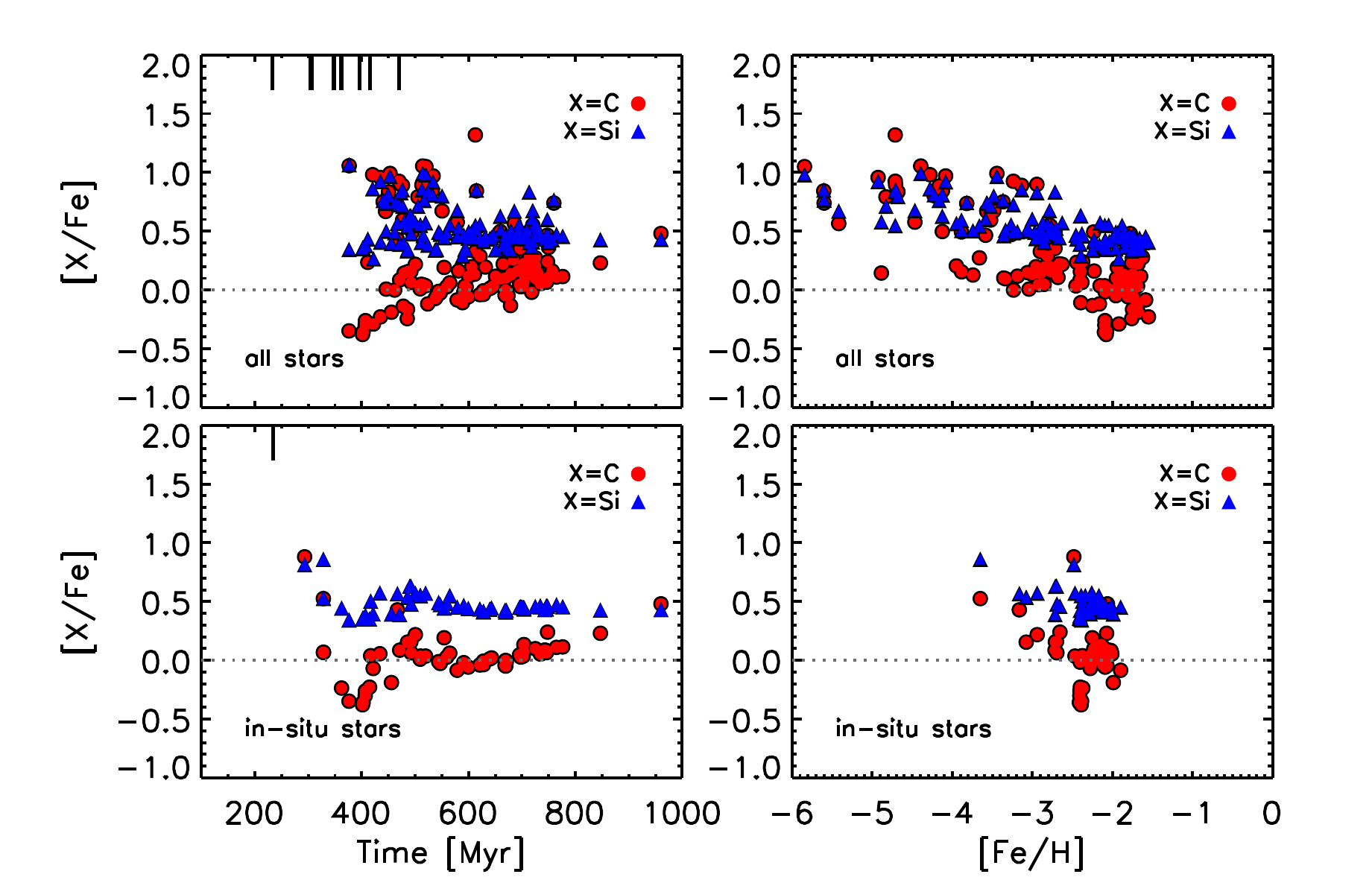}%{sedov_test.eps}
  \centering
   \caption{Comparison of abundance ratios, $\rm [C/Fe]$ and $\rm [\alpha/Fe]$, of all Pop~II stars in {\sc Halo1} ({\it top panels}) 
   versus in-situ stars ({\it bottom panels}). Following toy model results of Figure~7, we expect that Pop~III nucleosynthetic yields 
   show $\rm [\alpha/Fe]$ ratios that are similar 
   to $\rm [C/Fe]$, while $\rm [\alpha/Fe]$ ratios are higher than $\rm [C/Fe]$ in Pop~II SN yields. As such, in-situ 
   stars ({\it bottom}) appear to be predominantly born in environments contaminated by Pop~II SNe. 
   Among the {\it in-situ} stars, only a few stars, formed within a few 10 Myr after the first Pop~III SN 
   ({\it vertical mark on the top axis}), preserve Pop~III SN signatures. Low metallicity stars, with $\rm [Fe/H]\lesssim-4$, 
   exhibit the expected trend for Pop~III SNe, i.e. comparable $\rm [C/Fe]$ and $\rm [\alpha/Fe]$ ratios, implying that 
   they were formed out of gas enriched by Pop~III stars.}
\end{figure*}

\begin{figure*}
  \includegraphics[width=100mm]{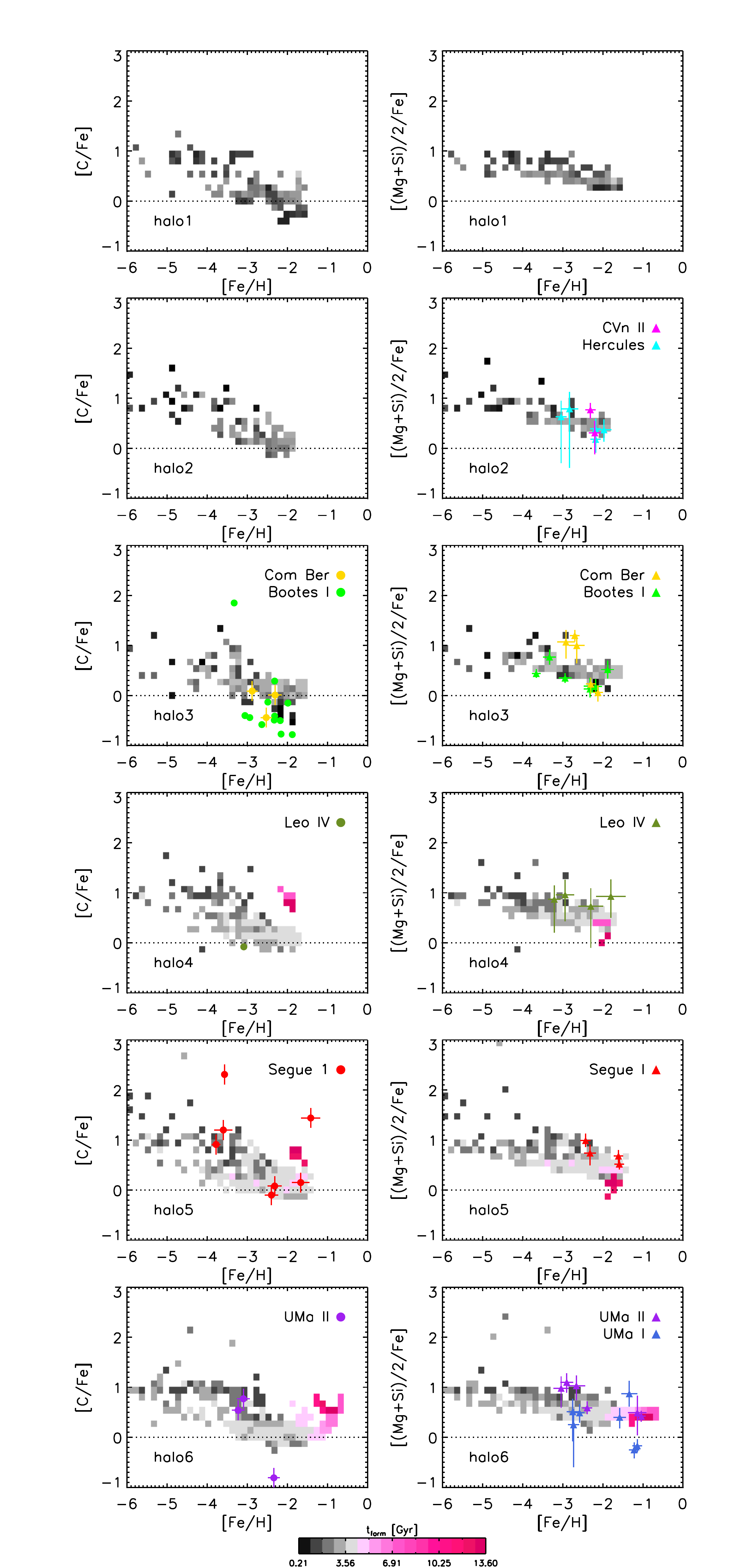}%{sedov_test.eps}
  \centering
   \caption{Stellar abundances in the simulated galaxies at $z=0$. {\it Left column:} $\rm [C/Fe]$ vs. $\rm [Fe/H]$. {\it Right column:} $\rm [\alpha/Fe]$ vs. $\rm [Fe/H]$.
The square symbols denote simulated abundances, with color indicating the formation
time of stars, such that grey colors mark stars formed
   before $t_{\rm H}\sim4$~Gyr, and reddish
   colors those formed after $t_{\rm H}\sim4$~Gyr. The existence of CEMP stars, particularly 
   at low metallicities, $\rm [Fe/H]\lesssim-3$, is a consequence of Pop~III SNe that intrinsically produce large amounts of carbon. 
   The enhanced $\alpha$-elements and the absence of clear knees, at which $\rm [\alpha/Fe]$ ratios decline, indicate that 
   the gas in the galaxies is preferentially metal-enriched by Type~II SNe, rather than Type~Ia, from Pop~III and massive Pop~II stars. In the 
   most massive {\sc Halo6} ({\it bottom panels}), the gas is metal-enriched up to $\rm [Fe/H]\sim-0.6$, owing to its extended star formation history. 
   Observational data of stellar abundances in select LG UFDs are displayed with the symbols, as labelled in the legends. For carbon, the data is taken from \citet{Simon2010} 
   for Leo~I, \citet{Frebel2010} for UMa~II and Com Ber, and \citet{Frebel2014} for Segue 1. For Bootes~I, the data is from 
   \citet{Frebel2016}, who provide a list of all high-resolution measurements from the literature. For the alpha-elements, we 
   use the data given by \citet{Frebel2010} and \citet{Gilmore2013}. A detailed comparison between the simulated UFD analogs 
and the observed UFDs is discussed in Section~3.5.}
\end{figure*}

\subsection{Chemical Abundances}
In this section, we first explain the general trends of the resultant chemical abundances, depending on 
different stellar populations, Pop~III and Pop~II. Then, we discuss individual abundances, carbon 
and alpha-elements based on the expected trends, and finally present metallicity distribution functions 
and comparisons to results for observed UFDs.
\subsubsection{General Trends}

To better understand our simulation results, we now introduce a simple toy model for the chemical evolution from multiple SNe. Specifically, in Figure~7 we illustrate the evolution of metal yields, 
as a function of time since the most massive star has exploded, for a Pop~III ({\it top}) and Pop~II ({\it bottom}) cluster. The panels on the left and right explore
different assumptions for the efficiency of metal transport (see below).
We only present the Pop~III yields up to $15$~Myr, 
which corresponds to the lifetime of the lowest-mass, and thus longest-lived, Pop~III star
for our chosen IMF, where $m_{\rm \ast, Pop~III}=10\msun$ (\citealp{Schaerer2002}). 
For Pop~II, we extend the yields to later times, corresponding to the
lower mass progenitor stars present in this case
($m_{\rm \ast, Pop~II}=8\msun$). 
In our idealized enrichment model, metals
are released in time such that those from the most 
massive stars are ejected first. Therefore, the gas metallicity at the SN explosion site changes 
over time, and one of the key factors is how rapidly metals can be transported
to large distances. For example, if the timescale for metal transport is shorter than 
the time between consecutive SNe, then the metallicity of a newly 
formed star will be determined mostly by SN yields from the most recent event.
In this case, metals from earlier SNe would
already have escaped into the surroundings. Conversely, if their propagating timescale 
is long, the metals ejected from different SNe will linger in the central region of 
a halo. The metallicity of a newly-formed star would then reflect averaged yields, 
deposited by multiple SNe. 
\par

To further construct our simple model, we consider 
two extreme cases of how efficiently metals are transported: (1) ejected 
metals are not accumulated in the central star forming region, and are instead immediately dispersed into the IGM; or (2) 
a certain fraction of the metals from each SN remains in the central region, thus giving rise to effective yields. The latter are calculated with a linear weight, measuring the time since explosion, such that the most recent SNe contribute
most to the average. Or formally, we employ a mass-dependent weight as
follows: $w(m_{\ast})=(t-\Delta t(m_{\ast}))/t$, where $t$ is the time
since the explosion of the most massive star in the cluster, and
$\Delta t(m_{\ast})$ is the time since a progenitor with mass $m_{\ast}$ exploded.
We find that the general trend of the yields from Pop~III and Pop~II SNe 
is substantially different. First, Pop~III SNe produce more carbon compared to Pop~II events. Second, 
for Pop~III yields, the $\rm [C/Fe]$ ratio is similar or larger than $\rm [\alpha/Fe]$, especially $\rm [Mg/Fe]$, 
while these trends are opposite for Pop~II yields.
These distinctive yield trends thus allow us to distinguish whether stars are formed under the dominant influence of Pop~III or Pop~II SNe. 
\par

We now compare these trends, obtained within the toy model above,
with the detailed enrichment results from our full simulations.
As discussed above, in-situ stars likely form in environments
internally enriched by Pop~II SNe, while stars formed
through external metal-enrichment are likely to show Pop~III signatures. 
These trends are evident in Figure~8, where we plot the chemical abundances of 
stars, formed in our simulations, as a function of their formation time ({\it left}) and  their metallicity ({\it right}). Recall that individual, 
reversed triangle symbols indicate single Pop~II clusters. Except for the very first stellar cluster, the majority of in-situ stars 
({\it bottom panels}) formed under the influence of Pop~II SNe, given that $[\rm Si/Fe]$ 
is larger than $\rm [C/Fe]$, which is the predicted trend for Pop~II SNe. As an exception, the very 
first stellar cluster was born out of gas enriched by Pop~III SNe, indicated by similar 
$\rm [C/Fe]$ and $\rm [Si/Fe]$ ratios, consistent with Pop~III SN yields.
The chemical abundances of all stars, shown in the upper panel of Figure~8, are established by a mixture 
of two systems: (1) haloes, mainly polluted via internal enrichment, hosting both Pop~III and Pop~II stars;
and (2) externally contaminated haloes. The existence of stellar clusters that exhibit comparable
$\rm [C/Fe]$ and $\rm [Si/Fe]$ ratios implies that their stars are likely formed out of gas mostly 
enriched by Pop~III stars. To the contrary, a lower value of $\rm [C/Fe]$ relative to $\rm [Si/Fe]$ signals the dominant contribution from
Pop~II SNe.

\subsubsection{Carbon}
Figure~9 shows select chemical abundances of 
Pop~II stars within the virial volume of our simulated galaxies at $z=0$, specifically
$\rm [C/Fe]-[Fe/H]$ and $\rm [\alpha/Fe]-[Fe/H]$ (left and right columns).
The color scheme denotes the formation time 
of stars, where greyish colors correspond to times before $t_{\rm H}\sim4$~Gyr and reddish ones to more recent epochs. 
As we have discussed in the previous sections, star formation in the three haloes, {\sc Halo1, Halo2}, and {\sc Halo3}, 
is truncated close to reionization, and thus {\it all} their stars form before 1~Gyr, shown as grey-scale colors. The stars in the other three haloes that 
experience late bursts or continuous star formation span a much wider range of formation times, reflected in the presence of square symbols with reddish colors.

\par
%CEMP stars
%reasons - Pop~III or binary or fallback
%Segue 1 - 
%Bootes 1 - self-enrichment
High- and medium-resolution spectroscopic studies of individual member stars in UFDs have revealed that the fraction of carbon-enhanced metal poor (CEMP) stars 
increases with decreasing metallicities (see \citealp{Beers2005} for details; and also, e.g., \citealp{Aoki2007}; \citealp{Norris2013}; \citealp{Yong2013}). 
Usually, CEMP stars are defined as metal-poor stars with carbon-to-iron ratios above $\rm [C/Fe]=0.7$ at $\rm [Fe/H]\lesssim-2$ (e.g. \citealp{Aoki2007}). 
Several possible mechanisms for their origin have been suggested, including {\rm (i)} rapidly-rotating massive Pop~III stars ($m_{\ast}=40-120\msun$), capable of releasing large amounts of
CNO-enhanced material (e.g. \citealp{Meynet2006}), {\it (ii)} binary 
systems with a companion star that undergoes an AGB phase, thus transferring 
carbon onto a low mass, long-lived, metal-poor star, or {\it (iii)} Pop~III SNe ($m_{\ast}=10-40\msun$) with a low explosion energy. In the latter case, iron-peak elements are more likely to be locked up in the emerging central remnant, whereas those with lower atomic number, such 
as carbon, are easily ejected. In addition, such ejecta are more likely to rapidly 
fall back since low SN energies are inefficient to drive the gas out of a host halo (e.g. \citealp{Umeda2003}; \citealp{Iwamoto2005}; \citealp{Cooke2014}; \citealp{Salvadori2015}).
\par

Our simulation results indeed exhibit the presence of CEMP stars at low metallicities. Since we exclude the existence of binary systems, the occurrence 
of CEMP stars in this work is a consequence of Pop~III SNe that intrinsically produce abundant carbon (see, Figure~7). Also, we find that the fraction of CEMP 
stars is higher at low metallicities, $\rm [Fe/H]\lesssim-4$, such that this fraction declines from $>45\%$ at $\rm [Fe/H]\lesssim-4$ 
to $<20\%$ at $\rm [Fe/H]\lesssim-2$ for all haloes. As demonstrated in the right-hand panels of Figure~8, low metallicity,
$\rm [Fe/H]\lesssim-3$, stars with high carbon-ratios of $\rm [C/Fe]>0.7$ originate from Pop~III SNe, characterized by their large intrinsic carbon production.
As can also be seen, the $\rm [C/Fe]$ ratio decreases toward near-solar values, giving rise to C-normal stars, with increasing metallicity. This behavior is expected as a consequence of
self-enrichment by Pop~II stars, where carbon yields are normal and the overall gas metallicity rises in the process.
\par

We now compare our simulated abundances with the empirical record in select UFDs. For instance, the member stars of Segue~1 roughly trace the trend of increasing number of CEMP stars
toward lower metallicity, except one star
with $[\rm C/Fe]=1.4$ at $\rm [Fe/H]=-1.6$.
Based on its enhanced abundance of neutron-capture elements,
\citet{Frebel2014} suggest that this outlier star might be associated with a binary system where
carbon is transferred from a companion. On the other hand, high-resolution spectroscopy shows 
that stars in Bootes~I both include CEMP signatures at low metallicities, 
($[\rm Fe/H]<-3.0$, and C-normal ones at $\rm [Fe/H]\gtrsim-3$ 
(e.g. \citealp{Gilmore2013}; \citealp{Ishigaki2014}; \citealp{Frebel2016}). Given the co-existence of CEMP and C-normal stars, Bootes~I thus appears to be
very similar to our simulated galaxies, in particular {\sc Halo3}, where both classes of stars are naturally produced through a combination of Pop~III SNe and self-enrichment by 
Pop~II stars.

%\begin{figure}
%  \includegraphics[width=80mm]{dwarf_info2.pdf}%{sedov_test.eps}
%   \caption{Stellar velocity dispersion as a function of V-band luminosity, comparing with other simulation results (\citealp{Simpson2013}), and select 
%   observations of UFDs (\citealp{Walker2009}).}
%\end{figure}

%Carbon enhancement is not required for very-low-iron abundance gas to cool and form low-mass stars.
%CEMP with neutron-capture s-process - binarity - carbon-rich due to the accretion of enriched material from an AGB companion.

\subsubsection{Alpha Elements}
%tells us about time scale
The run of $\rm [\alpha/Fe]$ over $\rm [Fe/H]$ serves as a well-known cosmic clock, encoding the timescale over which stars have formed. In particular, the existence of a knee,
where $\rm [\alpha/Fe]$ begins to decline, has been suggested as a reflection of a long 
duration of star formation. Given sufficient time, Type~Ia SNe would begin to contribute, producing more iron 
compared to alpha elements. As is evident in the right panels of Figure~9, we see roughly flat $\rm [\alpha/Fe]$ abundances at 
low metallicities, $\rm [Fe/H]\lesssim-3$, in all runs. As discussed in the previous sections, such 
low metallicity stars are likely to be formed via external metal enrichment by Pop~III SNe. We note that 
relatively massive haloes, {\sc Halo4, Halo5}, and {\sc Halo6}, show considerable scatter in the $\alpha$-abundance ratio, including values as high as $\rm [\alpha/Fe]\gtrsim2$. 
This scatter can be understood as follows. The number of progenitor haloes, hosting Pop~III stars, rises with increasing halo mass. 
As a result, the gas is incorporated from multiple haloes that are contaminated by different Pop~III SNe 
at a given time, leading to a large spread in $\rm [\alpha/Fe]$ ratio (see, Figure~7). We should mention that our 
simulations can only trace a few alpha elements, including Si and Mg, limiting our ability
to compare to observations.
\par

Among the simulations, the less massive haloes, {\sc Halo1, Halo2}, and {\sc Halo3}, exhibit enhanced $\rm [\alpha/Fe]$ ratios over the entire 
metallicity range, extending to $\rm [Fe/H]\sim-2$, without an obvious decline at high $\rm [Fe/H]$. The absence of a distinct knee 
implies that the stars in these haloes 
have formed out of gas that was preferentially polluted with metals from Type~II rather than Type~Ia SNe. In contrast, the most massive systems,  
{\sc Halo5} and {\sc Halo6}, show hints of Type~Ia SNe. For instance, stars formed late in {\sc Halo5} at $z\sim0.5$ (see Figure~2) have near- or super-solar
$\rm [\alpha/Fe]$ ratios at $\rm [Fe/H]\sim-1.6$. Unlike the other haloes, owing to the extended star formation, stellar metallicity 
in {\sc Halo6} increases extending to $\rm [Fe/H]\sim-0.6$. It should be noted that the $\rm [\alpha/Fe]$ ratio
at $\rm[Fe/H]\gtrsim-2$ is below that at lower metallicities, but $\rm [\alpha/Fe]$ still remains super-solar. This is due to the continuous star formation, followed 
by Type~II SNe, which are likely to wash out the impact of Type~Ia SNe, by continuously producing abundant alpha elements.
\par

 \begin{figure*}
  \includegraphics[width=130mm]{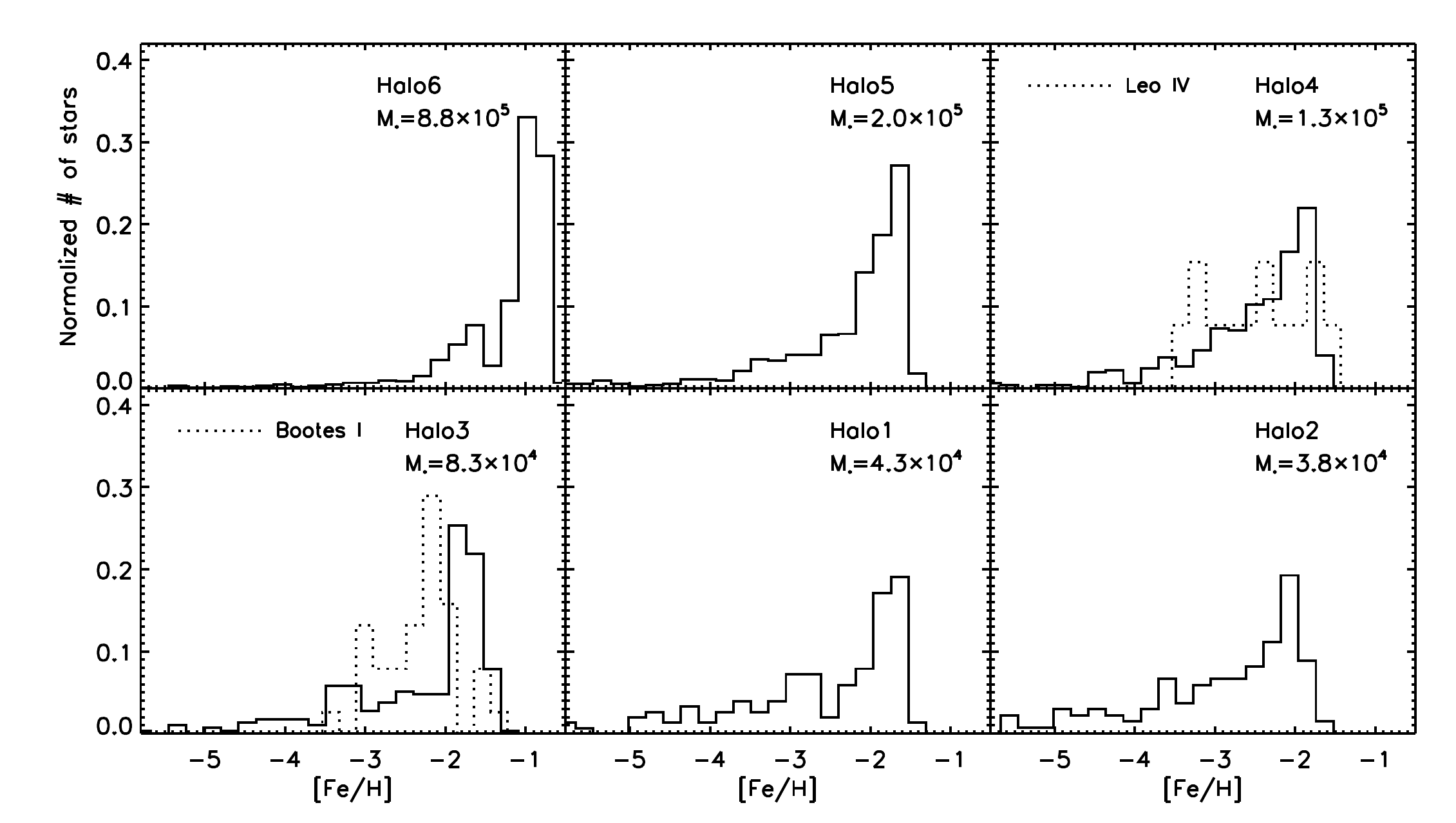}%{sedov_test.pdf}
  \centering
   \caption{Normalized metallicity distribution function (MDF) of Pop~II stars for each halo, ordered according to decreasing stellar mass (beginning in the upper-left panel). The MDF of haloes with stellar mass larger than $M_{\ast}>10^4\msun (z=0)$ shows a peak at [Fe/H]$\sim-2$, below which it gradually declines toward the low-metallicity regime. Such MDF shape is a natural consequence of both Pop~III and Pop~II SNe contributing: the peak at [Fe/H]$\sim-2$ is due to the continuing self-enrichment by Pop~II stars, whereas the low-metallicity tails represent stars formed in neighboring haloes under the influence of Pop~III SNe. As a result, such low-metallicity stars are likely to preserve the signatures of Pop~III SNe. Owing to the prolonged SFH of {\sc Halo6}, the MDF peaks at $\rm [Fe/H]\sim -1$, but none of the observed UFDs can explain such distribution. As explained, it is reasonable for {\sc Halo6} to be classified as low-mass dwarfs rather than UFDs. Alternatively possible explanation would be that environmental effects within the virial radius of the MW could play a critical role in terminating star formation in such systems.}
% Otherwise, they would have continuously formed stars in the field. - self-quenching}  
  %For low stellar-mass galaxies, $M_{\ast}<10^4\msun (z=0)$, the MDF is more evenly spread to low metallicity, showing a nearly flat distribution. This corresponds to systems in which only a few stars are formed by external metal enrichment, and where subsequent star formation is self-terminated by SN feedback, evacuating
   %all remaining gas. Afterwards, these systems are unable to replenish their gas supply, owing to their shallow potential well.
\end{figure*}

As a limiting case, \citet{Frebel2012} proposed 
a ``one-shot enrichment" scenario, where feedback from massive Pop~II stars, themselves showing pure Pop~III chemical signatures, evacuated the remaining gas from their
host halo, thus completely suppressing further star formation until $z=0$. Thus, any long-lived Pop~II stars in these systems
would exhibit a constant alpha-element ratio, $[\alpha/\rm Fe]=0.35$, which is solely determined by the yields 
of Pop~III Type~II SNe. This should be, however, considered an extreme case, valid for small, isolated haloes, not 
undergoing continued accretion or mergers. Thus, our simulated galaxies are not the case in which a halo grows in mass, making 
the gas in the halo with a deepened potential well harder to escape, rendering the effectiveness of stellar feedback weaker. 
Furthermore, the evacuated gas will begin to fall back within the order of a few 10~Myr, triggering subsequent episodes of star 
formation (e.g. \citealp{Ritter2012}; \citealp{Jeon2014}). This multi-generational star formation is reflected as scatter in 
the $[\alpha/ \rm Fe]$ ratios of observed LG dwarfs. 
\par

Measurements of $\alpha$-elements in MW UFDs are provided by several studies including \citet{Vargas2013}, where 
individual alpha abundance ratios are presented for 61 red giant branch stars in 8 UFDs. They found that all UFDs 
are likely to show high $\rm [\alpha/Fe]$ ratio at low $\rm [Fe/H]<-2.5$, which is consistent with the results of our simulations. 
In particular, Segue 1 and UMa~II exhibit enhanced alpha-abundances across the observed metallicity range 
$\rm -3.5<[Fe/H]<-1.0$, indicative of the signature of Type~II SNe, while the other six systems show a general trend of decreasing 
$\rm [\alpha/Fe]$ ratio with increasing $\rm [Fe/H]$. However, none of the observed UFDs has yet revealed $\rm [\alpha/Fe]$ ratios below 
$\rm [Fe/H]\sim-3.5$, preventing us from comparing our predictions on extremely low-metallicity stars with observations.
\par

%\citet{Vargas2013} find that five LG UFDs 
%(Coma Berenices, Canes Venatici~II, Ursa Major~I, Leo~IV and Hercules) exhibit the trend of increasing $[\alpha/\rm Fe]$ ratio with
%decreasing $\rm[Fe/H]$, and that two UFDs, Segue 1 and Ursa Major~II, show a constant, enhanced $[\alpha/\rm Fe]$ ratio
%over all metallicities extending to $\rm [Fe / H]\sim-1.0$, indicating a high star formation efficiency.

\par
Overall, our predictions for the $\rm [\alpha/Fe]$ ratio show good agreement with observations over the metallicity range 
$\rm -3.5<[Fe/H]<-1.0$. The observations also show that most UFDs exhibit no clear knee, favoring the interpretation
that their stars were born out of gas dominantly polluted by Type~II SNe (e.g. \citealp{Vargas2013}; \citealp{Gilmore2013}). 
To be specific, the measured $\rm [\alpha/Fe]$ ratios of CVn~II and Hercules are consistent 
with those of {\sc Halo2}, while the values of Coma Berenices and Bootes~I agree with those of {\sc Halo3}. These are 
relatively less massive haloes, $M_{\rm vir}\lesssim3\times10^9\msun (z=0)$,
experiencing truncated star formation. Furthermore, {\sc Halo4} and {\sc Halo5} show excellent agreement 
with Leo~IV and Segue~1, in combination with their measured $\rm [C/Fe]$ ratios and the observed maximum metallicity.

%Any noticeable decrease with [Fe/H] and are roughly constant at +0.6. Seg 1 shows enhanced abundance ratios even up to [Fe/H] ~ 1.6, 
%suggestive of a lack of pollution by Type Ia SNe.

%also appear higher toward lower [Fe/H]
%reported enhanced alpha abundances at [Fe/H]<-2.5.
%flat abundance patterns.
%In combination with their .. measurement,

\subsubsection{Metallicity Distribution Function}
In Figure~10, we display normalized stellar metallicity distribution functions (MDFs) for the simulated 
galaxies, i.e., the fraction of Pop~II stars as a function of $\rm [Fe/H]$, in order of decreasing 
stellar mass, beginning from the top-left panel. The general shape of the MDF exhibits a peak that 
gradually declines toward the low-metallicity regime, and sharply drops 
above the peak. The location of the peak shifts from $\rm [Fe/H]\sim-1$ for the most massive 
halo {\sc Halo6}, with a stellar mass of $\rm M_{\ast}\sim 8.8\times10^5\msun (z=0)$, to 
$\rm [Fe/H]\sim-2$ for {\sc Halo2}, where $\rm M_{\ast}\sim 3.8\times10^4\msun (z=0)$. 
Our derived MDFs reflect the combined action of external metal-enrichment by Pop~III and self-enrichment by 
Pop~II stars. Since the low-metallicity stars have predominantly originated in haloes affected by Pop~III SNe, 
the resultant fraction is expected to be small due to the brief Pop~III era, thus giving rise to the extended tails. 
The self-enrichment by Pop~II stars, on the other hand, increases the gas metallicity in their host halo, 
imprinting the peak at high $\rm [Fe/H]$. 
\par

%Intriguingly, flat MDFs are found in low-mass galaxies in which stars are mainly formed via external metal enrichment and where
%star formation is terminated prior to the onset of self-enrichment. In these systems with their shallow potential wells, SN feedback
%evacuates all available gas, without the ability to replenish their supply later on, thus effectively self-terminating. 
We also plot observed MDFs for select UFDs, shown as dotted lines, provided by \citet{Brown2014}. 
Our simulations can approximately reproduce the range of metallicity, especially the upper limit, for Leo~IV and Bootes~I, but fail to 
simultaneously account for the peak of the observed MDFs. We find that none of the MDFs of the observed UFDs is matched with 
the MDF of {\sc Halo6} that reflects the prolonged star formation down to $z=0$, resulting in the peak at 
the relatively high metallicity ($\rm [Fe/H]\gtrsim-1$) via self-enrichment. This discrepancy supports the 
assertion that the environmental effects within the virial radius of the MW are responsible for the truncated star 
formation in the relatively massive UFDs at late time, providing a constraint on the accretion epoch of UFDs. 
However, we should emphasize that the empirical record, in particular for low metallicity stars, is currently still 
very much subject to small-number statistics, such that definitive conclusions cannot be reached yet.

\subsection{Connection to the Observed UFDs.}
In this section, we attempt to provide insight into the quenching and accretion timescales of the observed 
UFDs around the MW based on the similarities of their chemical abundances to our simulated haloes. To be specific, we match 
the simulated UFD analogs with the observed LG UFDs 
by comparing the chemical properties, such as $\rm [\alpha/Fe]$, $\rm [C/Fe]$, 
and $\rm [Fe/H]_{\rm max}$ which is the maximum metallicity. A summary of this comparison is listed in Table~3. 
We note, that a comprehensive comparison should include other properties, such as the full SFH and gas 
content of each dwarf. However, we focus here on the impact of the new physics we have introduced, which manifests 
in the chemical abundances of the simulated haloes. As such, we include low-mass dwarf analogs, like {\sc Halo6}, in this comparison.
For $\rm [Fe/H]$ values, we only compare stars with metallicity higher than $\rm [Fe/H]\sim-3.5$, below which 
the observation could be incomplete due to the lack of ability of observing metal-poor stars with current observations. 
\par
As shown in Figure~9, we find that the maximum metallicities of the simulated galaxies, 
$\rm [Fe/H]_{\rm max}\sim-1.8$ ({\sc Halo2}) and $\rm [Fe/H]_{\rm max}\sim-1.5$ ({\sc Halo3}), 
are likely to be consistent with the most metal-rich stars among the observed UFDs, such as CVn~II, Leo~IV, Coma Berenices, 
Bootes~I, and Hercules. The absence of high metallicity stars above $\rm [Fe/H]\sim-1.5$ in such UFDs, in combination with the enhanced 
$\rm [\alpha/Fe]$ ratios, could imply that these UFDs might have experienced similar SFHs of the two galaxies, {\sc Halo2}, and {\sc Halo3}, that 
show the early quenching of star formation by reionization. This interpretation is in line with that of \citet{Brown2014}, where they 
report that five UFDs, CVn~II, Leo~IV, Bootes~I, Hercules, 
and CVn~II, show early truncated star formation based on the old ages of their stellar populations, finding 
that 80$\%$ of their stars formed prior to $z=6$. Here, we have demonstrated for the first time that 
the theoretically expected chemical evolution of UFD analogs formed in a cosmological framework are consistent 
with this early truncation of star formation.
\par
In addition to the excellent agreement of $\rm [\alpha/Fe]$ ratios of Bootes~I with those of {\sc Halo3}, the co-existence of CEMP 
and C-normal stars can be reproduced in {\sc Halo3}, in which CEMPs are likely to be formed via Pop~III SNe, while C-normal 
stars are a consequence of self-enrichment by Pop~II stars. On the other hand, Segue~1 appears to be similar to {\sc Halo4}, 
providing good agreement between the observed 
and simulated chemical properties such as $\rm [Fe/H]_{\rm max}\sim-1.42$, $\rm [C/Fe]$, and $\rm [\alpha/Fe]$. 
As pointed out in Section 3.4.2, the star with $\rm [C/Fe]=1.4$ at $\rm [Fe/H]=-1.6$ might originate from a binary 
system (\citealp{Frebel2014}). Therefore, we could ignore this outlier since binary systems are not considered in this work. 
We find that {\sc Halo4} exhibits bursty star formation at late epochs at $z\sim3.9$, 
during which time stars are formed with relatively high metallicities ($\rm [Fe/H]\sim-1.6$). This suggests that star formation in 
Segue~1 might be not completely truncated by reionization, but followed by late bursts of star formation that gives rise to 
stars with $\rm [Fe/H]\gtrsim-1.6$. We find that none of the simulated UFD analogs exhibit similar chemical properties of UMa~I.
\par
%These simulations can thus be used to predict the SFH for a given UFD, based on its observed chemical 
%properties. These simulations can provide critical guidance on how to translate observed abundance patterns into a SFH. 

For example, the $\rm [\alpha/Fe]$ ratios 
of UMa~II are consistent with the derived $\rm [\alpha/Fe]$ ratios of {\sc Halo6}, which is a halo that shows 
extended star formation beyond reionization. However, {\sc Halo6} fails to match $\rm [Fe/H]_{\rm max}=-1$ 
of UMa~II (\citealp{Gilmore2013}), by producing stars with metallicity higher than $\rm [Fe/H]\sim-1$. 
This discrepancy might indicate that reionization was insufficient to 
completely suppress star formation in UMa~II, allowing it to continuously form stars after reionization, but afterwards 
it is clear that star formation was truncated by other factors, i.e., gas stripping during the infall into the MW halo. 
The simulations suggest that without this truncation UMa~II might have continuously or episodically formed stars 
down to $z=0$, producing stars with metallicity above $\rm[Fe/H]\sim-1$, as predicted in {\sc Halo6}.
\par
If we assume that the end of star formation of UMa~II is instead correlated with the infall time into the MW halo, the simulated 
analog, {\sc Halo6}, allows us to estimate the infall time of UMa~II. In Figure~11, we present the evolution of chemical 
abundances of {\sc Halo6}, $\rm [\alpha/Fe]$ vs. $\rm [Fe/H]$, at different 
redshifts, $z=7$ (top), $z=6$ (middle), and $z=3$ (bottom), with those of UMa~II at $z=0$. As shown in 
the middle panel of Figure~11, the larger value of $\rm [Fe/H]_{\rm max}\sim-1$ of the most metal-rich stars in UMa~II, compared to 
that of {\sc Halo6}, indicates an incapability of complete quenching of star formation by reionization. As stars in {\sc Halo6} 
are continuously formed, the metallicity increases, eventually forming stars with metallicity as high as $\rm[Fe/H]\sim-1$ 
at $z\sim3$. Provided that there are no observed stars with metallicity higher than $\rm[Fe/H]\sim-1$ in UMa~II, our 
simulations indicate that its SFH was truncated at $z\sim3$.
\par
If we assume that the infall onto the MW was responsible for the quenching of star formation in UMa~II, 
we can estimate that UMa~II fell in $\sim11$ Gyr ago. This prediction is consistent with the results of \citet{Rocha2012}, where 
they investigate the infall times for MW dwarfs using their present-day kinematics and the energy-infall relation. 
They suggest that UMa~II tends to support an early infall with $t_{\rm infall}\sim8-11$ Gyr. 
\par
We note that it is impossible to infer the accurate infall times 
of the observed UFDs from this work, since the simulated galaxies are in the field, which is beyond the virial radius of the MW-like host halo. 
Furthermore, based on the on-going star formation and the substantial amount of gas reservoir at $z=0$, {\sc Halo6} should be
considered as a low-mass dwarf rather than UFDs. However, this simulation could provide useful guidance on estimating the time when SFH of UFDs is 
truncated based on the maximum stellar metallicity observed.

\par

 \begin{figure}
  \includegraphics[width=75mm]{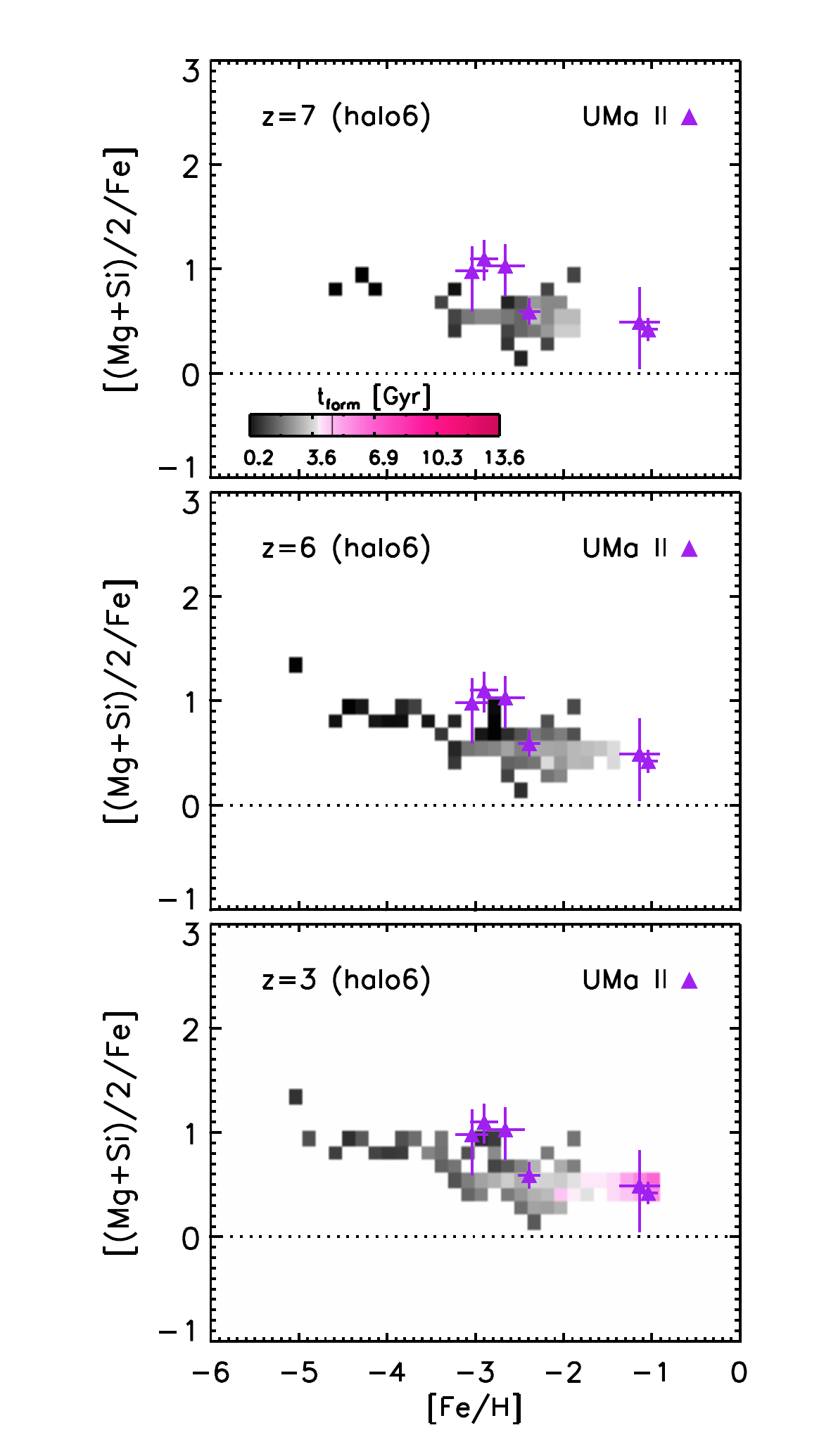}%{sedov_test.eps}
   \caption{The evolution of stellar abundances, [$\rm \alpha/Fe$] vs. [Fe/H], in {\sc Halo 6} at different redshifts, $z=7$ ({\it Top}), $z=6$ ({\it Middle}), \rm and $z=3$ ({\it Bottom}), by comparing with the observed values of UMa~II at $z=0$ (\citealp{Gilmore2013}). As demonstrated in the middle panel, the discrepancy in 
   $\rm [Fe/H]_{\rm max}\sim-1$ between {\sc Halo6} and UMa~II could indicate that reionization was insufficient to completely quench star formation in UMa~II. {\sc Halo6} forms stars with such metallicities at $z=3$ ({\it bottom panel}), meaning that UMa~II formed stars down to $z=3$ and truncated afterwards. Furthermore, if we assume that the end of star formation of UMa~II is correlated with gas stripping while falling onto the MW halo, then the {\sc Halo6} results suggest an estimated infall time for UMa~II that is as old as $\sim$11 Gyr ago.} 
 
\end{figure}

\begin{figure}
  \includegraphics[width=75mm]{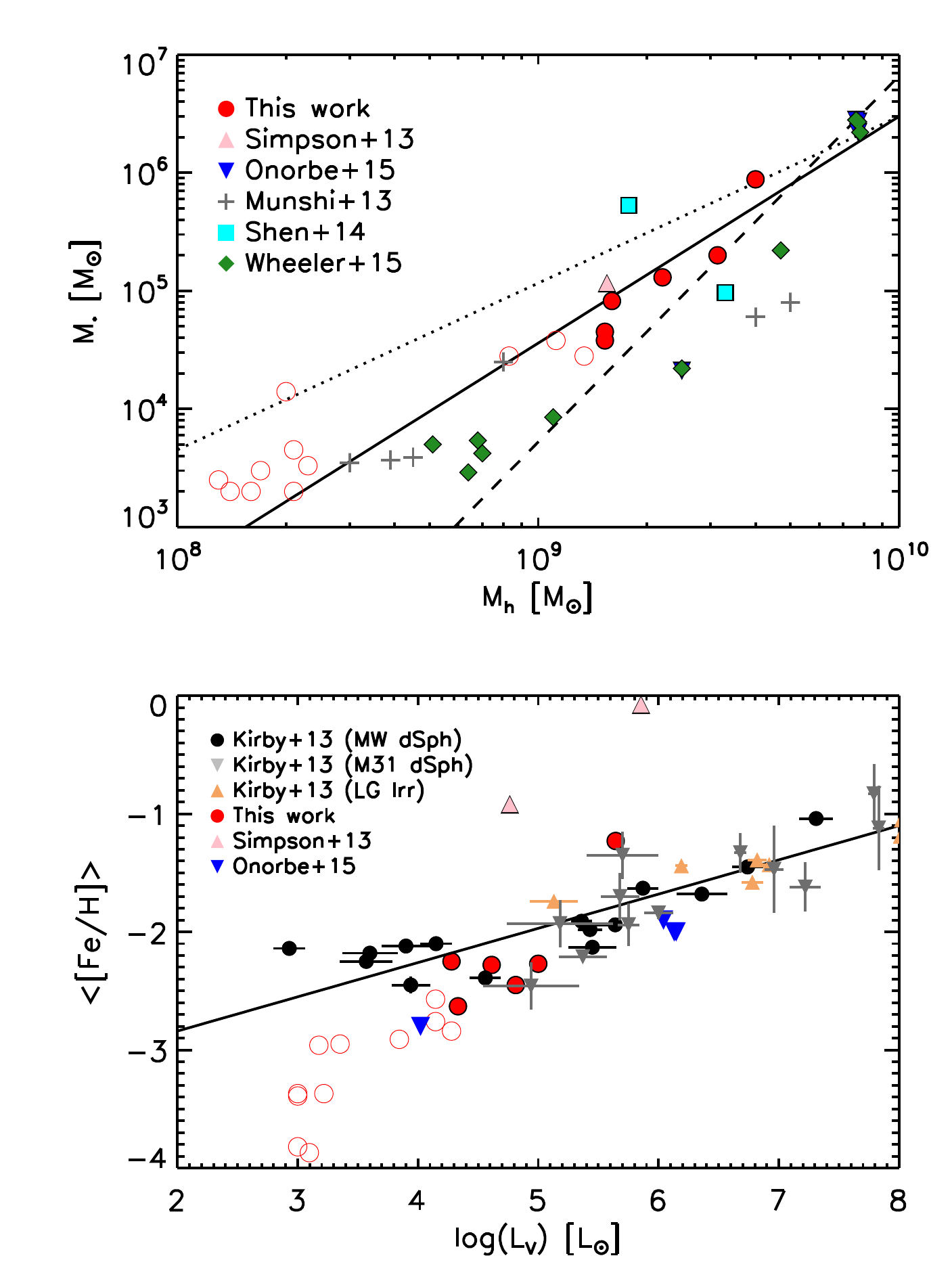}%{sedov_test.eps}
   \caption{Global physical properties of the simulated dwarf galaxies at $z=0$, stellar mass vs. halo mass {(\it top)} and stellar metallicity vs. luminosity {(\it bottom)}. 
   The quantities derived from our work are shown as red filled circles. We also show results from other hydrodynamic simulations 
   (\citealp{Simpson2013}; \citealp{Munshi2013}; \citealp{Shen2014}; \citealp{Wheeler2015}; and \citealp{Onorbe2015}), as well as select observations 
   (\citealp{Kirby2013}). In the top panel, we compare our stellar-halo mass results to theoretical fits from abundance matching techniques 
   (\citealp{Behroozi2013} (dotted line), \citealp{Brook2014} (dashed line), and \citealp{GK2016} (solid line)); as can be seen, our results show good agreement 
   with the fit suggested by \citet{GK2016}. We also mark the properties of low-mass galaxies ($M_{\rm halo}<10^9\msun (z=0)$, red open circles) 
   that are located around the six galaxies ({\sc Halo1-Halo6}). These galaxies are composed of only a small number of stellar clusters formed via external 
   metal-enrichment, giving rise to low average metallicity.} 
\end{figure}

\begin{deluxetable}{c c c c}
%\centering
\tablecolumns{4}
\tablewidth{0.4\textwidth}
\tablecaption{Summary of comparison between the simulated haloes and the abundances of observed UFDs.}
\tablehead{\colhead{UFDs} & \colhead{$M_{\ast}[\msun]$} & \colhead{Analogs} & \colhead{Similar properties}}
\startdata
\vspace{0.07cm}
{Segue~I} &$1.0\times10^3$ & {\sc Halo5}& $\rm [\alpha/Fe]$, $\rm [C/Fe]$, $\rm [Fe/H]_{\rm max}$ \\ 
\vspace{0.07cm}
{Com Ber} & $3.7\times10^3$ & {\sc Halo3} & $\rm [\alpha/Fe]$, $\rm [C/Fe]$  \\
\vspace{0.07cm}
{UMa~II} & $4.1\times10^3$ & {\sc Halo6}& $\rm [\alpha/Fe]$, $\rm [Fe/H]_{\rm max}$\\
\vspace{0.07cm}
{CVn~II} &  $7.9\times10^3$& {\sc Halo2} & $\rm [\alpha/Fe]$ \\
\vspace{0.07cm}
{Leo~IV} & $1.0\times10^4$ &{\sc Halo4} & $\rm [\alpha/Fe]$, $\rm [C/Fe]$, $\rm [Fe/H]_{\rm max}$\\
\vspace{0.07cm}
{UMa~I} &  $1.4\times10^4$& {\sc None} & - \\
\vspace{0.07cm}
{Bootes~I} &$2.9\times10^4$& {\sc Halo3} & $\rm [\alpha/Fe]$, $\rm [C/Fe]$, $\rm [Fe/H]_{\rm max}$\\
\vspace{0.07cm}
{Hercules} &$3.7\times10^4$& {\sc Halo2}  &  $\rm [\alpha/Fe]$, $\rm [C/Fe]$, $\rm [Fe/H]_{\rm max}$\\
\enddata
\vspace{-0.1cm}
\tablecomments{Column (1): the names of the observed UFDs. Column (2): stellar mass of the observed UFDs in $\msun$. 
Column (3): the name of the most similar analogs among the simulated galaxies based on the 
chemical properties listed in Column 4.}
\end{deluxetable}

\subsection{Global Galaxy Properties}
We present global galaxy properties at $z=0$ inferred from the simulations in Figure~12. The relationship 
between stellar and halo mass of galaxies is shown in the top panel, where the red filled circles 
denote our simulations. Results from other works are also displayed, including hydrodynamic
simulations (e.g. \citealp{Simpson2013}; 
\citealp{Munshi2013}; \citealp{Shen2014}; \citealp{Onorbe2015}; \citealp{Wheeler2015}), and
abundance matching using dark-matter simulations (e.g. \citealp{Behroozi2013}; \citealp{Brook2014}; \citealp{GK2016}). 
Given that our spatial and mass resolutions are comparable to \citet{Wheeler2015} and 
\citet{Onorbe2015}, the discrepancy in the predicted stellar masses might be attributed to
additional feedback effects considered there. Specifically, they also take into account 
the impact of photoheating and radiation pressure from stars, which could reduce star formation.
\par
Our results show good agreement with a fit, $M_{\rm \ast}\propto M_{\rm vir}^{1.8}$,
suggested by \citet{GK2016}, who derived the relation based on LG galaxy counts and 
the results of dark-matter simulations.
The average stellar metallicity vs. V-band luminosity is presented in the bottom panel 
of Figure~12, and compared again with other works and observations.
The V-band luminosity is computed by assuming a stellar mass-to-light ratio of 2$\msun/\lsun$ for an old 
stellar population (\citealp{Kruijssen2009}). The average stellar metallicity ranges from $\rm [Fe/H]=-1.23$ 
for a few $5\times10^5\lsun$ galaxy to $\rm [Fe/H]=-2.63$ for $2\times10^4\lsun$ systems, which is in good agreement with observations.

%We find that the galaxy sizes, described as half stellar mass radii, are similar 
%to the estimates from other hydrodynamic simulations, while predictions from the simulations are, overall, larger
%than to those inferred from observations of LG UFDs (e.g. \citealp{Wolf2010}).
\par
In addition to the six haloes ({\sc Halo1-Halo6}), we also present the properties of small galaxies 
($M_{\rm halo}<10^9\msun (z=0)$, red open circles) around the six haloes in the refined regions. 
A noticeable discrepancy from the observed relation is that 
such small galaxies result in much smaller stellar metallicity, by an order of magnitude, at 
luminosities below $L_{\rm V}\sim10^{3.5}\lsun$. This corresponds to systems in which only a 
few stars are formed by external metal enrichment, and where subsequent star formation is 
self-terminated by SN feedback, evacuating all remaining gas. Afterwards, these systems are 
unable to replenish their gas supply, owing to their shallow potential well. As a result, those 
low-mass galaxies are composed of only a small number of stellar clusters that form in 
externally-enriched environments, giving rise to low average metallicity. However, the robustness 
of the derived properties of such small galaxies is subject to mass resolution. Therefore, we plan to firm up 
this possible external enrichment pathway with even higher-resolution simulations in future work.

%The current lack of observations 
%of such extremely low-metallicity stellar populations might be due to an observational bias toward relatively 
%metal-rich stars.
%A final comparison in Figure~12 is the stellar velocity dispersion of the 
%simulated galaxies versus V-band luminosity. The derived velocity dispersion ranges from $\sim$ 2 km $\rm s^{-1}$ 
%for a $10^3\lsun$ galaxy to a few 10 km $\rm s^{-1}$ for $10^5\lsun$ systems, which is in good agreement with observations. 

\par

%We next present our results for the relations of V-band luminosity vs. half-stellar-mass radius
%and total mass within half-stellar-mass radius vs. the same radius
%(upper righ-hand and bottom left-hand panels, respectively), and compare again with other works and observations.
%The V-band luminosity is computed by assuming a stellar mass-to-light ratio of 2$\msun/\lsun$ for an old 
%stellar population (\citealp{Kruijssen2009}). We find that the galaxy sizes, described as half stellar mass radii, are similar 
%to the estimates from other hydrodynamic simulations, while predictions from the simulations are, overall, larger
%than to those inferred from observations of LG UFDs (e.g. \citealp{Wolf2010}).
%\par

%In order to assess the observability of the simulated 
%galaxies, we additionally plot the surface brightness detection limits of 30 mag arcsec$^{-2}$ and 32.5 mag arcsec$^{-2}$ for the 
%Sloan Digital Sky Survey (SDSS, dotted line) and the Large Synoptic Survey Telescope (LSST, solid line), respectively. 
%All the simulated galaxies, except two that show relatively larger galaxy size at a given luminosity, are above the detection limit 
%of the upcoming co-added LSST. However, since the co-added LSST will be able to search for objects as faint as $\sim2\times10^3\lsun$ 
%out to distances of $\sim1$ Mpc (see fig.~9 in \citealp{Tollerud2008}), only two out of the six zoom-in regions, located within 1 Mpc 
%from the primary halo, will be within reach.
\par

\section{Caveats and Limitations}
In this section, we briefly comment on the limitations of this work and our adopted assumptions that could affect the results.
One of the robust results of our work is the key role of external metal enrichment in the formation of extremely metal-poor 
stars. The degree of external metal enrichment, however, heavily depends on how we implement the transport of metals and 
their mixing with the surrounding gas. In the diffusion method, adopted here, the efficiency of mixing is determined by the diffusion coefficient 
described in Section~2.4.3.
\par
To test the robustness of our results regarding the efficiency of metal mixing, we additionally 
perform a comparison simulation. Here, we artificially reduce the diffusion coefficient by a factor of 5 in the {\sc Halo1} zoom-simulation, and 
examine if we still can produce extremely low metallicity stars. We find that the fraction of stars, 
defined as $f =$ (number of stars with metallicity less than a given $\rm [Fe/H]$) / (total number of stars), decreases for 
$f(\rm [Fe/H]<-3)$, $f(\rm [Fe/H<-4])$, and $f(\rm [Fe/H]<-5)$ by a factor of 1.5, 1.2, 1.04, respectively. Given the stochastic nature of star formation, we conclude that the
ability of producing low-metallicity stars within our simulations is not very sensitive to changes in the diffusion coefficient. 
We note that external metal-enrichment is argued to be much less important in the recent work by \citet{Griffen2016}, 
where they investigate the fossils of Pop~III star formation and of the first galaxies in a MW-mass system. 
The idealized model for metal pollution, implemented in post-processing within their dark-matter only simulation, however, makes it difficult
to directly compare with their results. 
\par
%Comparing test between diffusion and smoothed particle scheme on the metal mixing is discussed in \citet{Revaz2016} 
%in which they found that both methods can reproduce the scatter of the chemical abundances to a realistic value of the observed 
%dSphs.

Other uncertainty that could affect the resulting chemical abundances would be the IMF of Pop~III and 
Pop~II stars. As previously mentioned, the Pop~III IMF, in particular, is still very uncertain. Here, we only consider the mass range between 
$10\msun$ and $150\msun$ by assuming a top-heavy IMF, but recent high-resolution studies of Pop~III star formation 
have suggested that the mass of primordial stars could extend to as low as $m_{\ast}<1\msun$ (e.g. \citealp{Stacy2016}). 
If this were the case, our simulations would have overproduced metals, and overpredicted alpha-element abundances as well, via CCSNe and 
PISNe from massive Pop~III stars. In addition, the low SFRs in the simulated galaxies indicate that IMF sampling is likely to be incomplete. 
However, this stochastic IMF sampling is a reasonable assumption for Pop~III clusters. This is because Pop~III stars tend to be formed as single or multiple stars 
and the following energetic SNe prevent further star formation in small halos ($M_{\rm vir}\lesssim10^7\msun$ at $z>10$). Therefore, it is expected that 
the chemical abundances of Pop~II stars are predominately determined by the yields from one or a few Pop~III SNe.
\par

We do not take into account photoheating and radiation pressure from stars. For the low-mass galaxies simulated here, 
these additional stellar feedback effects may play an important role in suppressing star formation. In this context, \citet{Wise2012} 
have reported that the star formation rate can be reduced by a factor of five in a dwarf galaxy at $z\sim8$ with a
virial mass of $2\times10^8\msun$ by including radiation pressure. Similarly, the key role of
radiative feedback in predicting the stellar masses of galaxies, in particular at
high redshifts, is emphasized in the study
by \citet{Hopkins2013}.
However, it is currently still prohibitive computationally to trace the evolution of a galaxy
down to $z=0$, while directly solving the radiative transfer equation for our highly resolved star forming 
regions within their large-scale cosmological context.
\par

It should also be pointed out that we fix the onset of reionization by introducing an external UV background at $z=7$, ramping it up to its full strength 
by $z=6$. \citet{Simpson2013}, on the other hand, have shown that depending on the onset of reionization, the resultant stellar mass 
at $z=0$ can vary by an order of magnitude (see also \citealp{Milos2014}). The precise time when a given galaxy is impacted by the UV background
would depend on how far the galaxy is located from massive haloes, where the majority of UV photons is emitted, implying that 
dwarf galaxies close to a massive system might experience an early truncation of star formation. In addition, apart from the uniform 
UV background, local heating and ionization from the nearby galaxy are neglected. The effect of local radiation sources and the impact of the 
MW potential will be presented in a subsequent paper. Recently, \citet{Onorbe2017} have provided a new set of UV background to complement 
the extensively used heating and ionization rates (\citealp{Haardt2012}), which turn out to be far strong at early epochs ($z>6$), heating the temperature 
of the IGM to $10^4$ K already at $z=13$. It means that star formation in the simulated galaxies could have terminated at early times than it 
should, resulting in the overestimated stellar mass at $z=0$. However, we expect that our results may be insensitive to this difference because we introduce the UV background 
from $z=7$ where two models begin to produce similar reionization history and the IGM temperature.
\par

The final point is that the size of our simulation box is relatively small, dictated by the
need to keep computational cost under control. We cannot, therefore, properly represent the large-scale
modes in the cosmological density field that would control the number of structures at $z=0$, and would
also affect the accretion history of the simulated haloes. However, in this work, we focus on the evolution 
of individual systems rather than the statistics of sub-structures of a realistic MW-size halo. In addition,
performing simulations for multiple refined regions allows us to investigate the build-up of galaxies under
different accretion histories, thus partially addressing the impact of cosmic variance.

\section{Summary and Conclusions}
We have performed a suite of cosmological hydrodynamic zoom-in simulations to investigate 
the star formation history and chemical evolution of local low-mass dwarf galaxies. 
We have, for the first time, traced the detailed enrichment history of stellar populations in 
UFD galaxies, reaching back to the formation of Pop~III stars. Our work, therefore, allows us 
to directly compare the derived stellar chemical abundances, including the contribution from
the first stars, to the detailed observations of individual stellar abundances in UFDs. %Specifically, we
%have traced their evolution over the entire age of the Universe, by following the chemical 
%signatures of Population~III (Pop~III) stars for the first time. This allows us to directly compare our
%simulated galaxies to their proposed descendant fossil population, local ultra faint dwarfs (UFDs). 
Given the physical characteristics of the local UFDs, which consist of metal-poor, ancient 
stellar populations ($\gtrsim$ 10 Gyr), there have been theoretical attempts to link the nearby UFDs 
to galaxies at high redshifts ($z\gtrsim6$). In particular, the truncated star formation history 
of UFDs, inferred from observations, supports the idea that 
their star formation could be partially or completely quenched by reionization. The gas
in dwarf galaxies was then photo-heated, and the filtering (Jeans) mass in the reionized IGM was increased, such that
gas collapse into low-mass haloes became inefficient.
\par

Our simulations confirm that reionization, in combination with supernova feedback, did play a critical role 
in suppressing star formation in local dwarf
galaxies. The effectiveness of reionization in preventing star formation strongly depends on halo mass: 
star formation in low-mass haloes, with $M_{\rm vir}\lesssim 2\times10^9\msun (z=0)$, 
is almost entirely quenched by reionization, while relatively massive haloes, $M_{\rm vir} \gtrsim 3\times10^9\msun (z=0)$, 
exhibit a continuous or bursty star formation history at late epochs (e.g., our {\sc Halo6}). 
%Our results might imply a crucial role of environmental effects of the MW on quenching 
%star formation in the UFDs that are massive enough not to be entirely truncated by reionization. 
We find that in most cases, more than $90\%$ 
of stars form prior to reionization, such that these systems can be classified as ``true fossils". On the other hand, 
the most massive halo, {\sc Halo6}, with $\rm M_{\rm vir}\sim4\times10^9\msun (z=0)$, forms only about $30\%$ of stars before reionization, 
and star formation is not effectively shut down. Such a galaxy would correspond to a ``polluted fossil" in terms of the definition of 
\citet{Ricotti2005}.
The comparison run without SN feedback confirms that this feedback is essential in truncating
star formation in dwarf galaxies.
\par
The ability of resolving small haloes ($M_{\rm vir}<10^8\msun$ at $z>6$) enables us to track down stellar mass growth in minihalo 
scales at high-z, leading to very interesting results. Our simulations find that star formation in the simulated galaxies, especially 
less massive haloes ($M_{\rm vir}<2\times10^9\msun$ at $z=0$), is a consequence of the combination of in situ star formation in a primary 
halo and significant stellar accretion through multiple mergers of smaller haloes, that originate from low-density peaks. 
This conflicts with the prevailing view that stellar mass growth is dominated by in-situ SF within a dominant primary halo - instead, it is difficult to identify a primary halo at early times for UFD analogs. This suggests that when we observe stars in UFDs it is not simple to infer 
where the stars originate, complicating efforts to reconstruct their SFHs.
On the other hand, the most massive halo, {\sc Halo6}, 
presents the expected trend that more than 80\% of stars are formed in a primary halo. This case is because the progenitor haloes of 
{\sc Halo6} were from high-density peaks and the primary halo grow fast through accretion and mergers. This suggests that there is a 
distinction between the assembly history of UFDs and low-mass dwarfs.

\par
Our simulations demonstrate that the inclusion of Pop~III SNe is a necessary ingredient to produce low metallicity stars with
$\rm [Fe/H]\lesssim-4$. Particularly, we find that the existence of such low-metallicity stars is predominantly a consequence of 
external metal enrichment by Pop~III SNe. Due to the proximity between haloes at high 
redshifts, $z\gtrsim7$, metals from Pop~III SNe can be transferred into neighboring haloes that have 
never formed stars, but are about to do so. As a result, extremely low-metallicity Pop~II stars are likely to be born in this 
externally contaminated environment.
Based on the distinctive trends of metal yields from Pop~III and Pop~II SNe, we scrutinize
the detailed origin of the stellar components in the dwarfs. We show that the simulated galaxies are 
composite systems, assembled from haloes that were externally enriched, hosting only Pop~II stars, and from
haloes in which the gas was polluted via self-enrichment, hosting both Pop~III and Pop~II star formation. Our simulations 
naturally reproduce the carbon-enhanced metal poor stars by including Pop~III SNe that intrinsically yield 
high ratios of $\rm [C/Fe]$. Also, the resultant chemical abundances provide a good match to the observations,
in that we find the enhanced $\rm [\alpha/Fe]$ ratios at all metallicities, a trend also found 
in the observations of UFDs. This enhanced $\rm [\alpha/Fe]$ ratio, nearly constant with metallicity, implies that the gas 
in the dwarfs is more likely affected by Type~II rather than Type~Ia SNe. 

%such that Pop~III stars produce 
%comparable $\rm [Si/Fe]$ ratios to $\rm [C/Fe]$, whereas $\rm [Si/Fe]$ ratios are higher than 
%$\rm [C/Fe]$ ratios from Pop~II SNe yields. 

%Finally, our simulations show a good match with other theoretical works and the 
%observed properties of dwarf galaxies.
\par

Comparing the derived chemical properties of the simulated galaxies with those of the observed UFDs 
(CVn~II, Leo~IV, Coma Berenices, Bootes~I, Hercules, Segue~I, UMa~I, and UMa~II), we find that 
all these UFDs appear to have experienced truncated star formation. Especially, the chemical properties of 
four UFDs, CVn~II, Hercules, Coma Berenices, and Bootes~I, tend to be consistent with those of 
the simulated analogs, {\sc Halo2} and {\sc Halo3}, possibly implying that 
star formation in these UFDs was entirely quenched by reionization. The presence of relatively high metallicity stars in Leo~IV and 
Segue~I might be associated with bursts of star formation at late epochs after reionization, as suggested in {\sc Halo4} and {\sc Halo5}.

The most massive dwarf halo in our sample, {\sc Halo6} (M$_{\rm vir} \sim 4\times10^9 \Msun$ at $z=0$), 
continues forming stars until z=0 and retains a significant neutral HI reservoir at z=0.  The amount of neutral hydrogen and 
stellar mass at z=0 is comparable to that of newly discovered gas-rich, low-mass dwarfs like Leo A, Leo P, Leo T and DDO 210. 
This work indicates that the progenitors of such 
gas-rich dwarfs are slightly more massive than true fossil relics of reionization. 
%The study of {\sc Halo6} further informs us about the origin 
%and evolution of UFDs with high metallicity tails. The similarity in the chemical abundance of {\sc Halo6} at z=3 to Uma~II suggests that it 
%was born in a higher mass halo than other UFDs that was then quenched at z$\sim$3. 
This study indicates for the first time the significance 
of the chemical evolution of low-mass haloes to z=0 in a full cosmological context to make direct comparison with the full variety of dependent
dwarfs observed around the MW.  In particular, 
%Furthermore, we estimate the time at which the most metal-rich stars of UMa~II were generated, which allows us to infer 
%the end of star formation in UMa~II at $z\sim3$.
if we assume that this quenching of star formation in UMa~II was correlated with 
gas stripping while infalling onto the MW halo, the estimated infall time of UMa~II could be $t_{\rm infall}\sim11$ Gyr.
However, {\sc Halo6} already looks very different from UFDs ($M_{\ast}<10^5\msun$ at z=0), forming $30\%$ of 
stars after reionization, giving rise to a stellar mass of $M_{\ast}=2.6\times10^5\msun$ ($z=6$), meaning that there likely is a hard upper limit 
on the UFD halo mass. It also suggests that UMa~I and UMa~II could be the most massive UFDs among the observed UFDs, but 
still not as massive as {\sc Halo6}.

\par

A key prediction of our simulations is the presence of extremely low-metallicity stars, with $\rm [Fe/H]\lesssim-4$, in the local UFDs. These stars should 
preserve pure Pop~III signatures, i.e. enhanced carbon and alpha abundances, as a consequence 
of external metal-enrichment and the build-up of dwarfs via mergers. 
To date, such extremely low-metallicity stars have not yet been detected in UFDs, but
they should come within reach over the next decade, with the advent of a number of
next-generation telescopes. Among them is the Large Synoptic Survey Telescope (LSST) with its
unprecedented wide-field imaging capability, ideally suited for detecting low surface-brightness systems. In addition, the greatly improved spectroscopic sensitivities
of the upcoming extremely large ground-based telescopes, the Giant Magellan Telescope (GMT),
the Thirty Meter Telescope (TMT), and the European Extremely Large Telescope (E-ELT), will
revolutionize chemical abundance studies in the Local Group. Also, a recent
study of the most metal-poor damped Lyman-$\alpha$ (DLA) system at $z\sim3$, with an iron
abundance of $\rm [Fe/H]\lesssim-2.81$, suggests that the chemical signatures 
from a core-collapse supernova of a $20.5\msun $Pop~III star are imprinted on the observed DLA, indicating that such systems
could be the predecessors of UFDs (\citealp{Cooke2017}). All these avenues promise to offer fundamental insights into the nature of the first generation 
of stars and into the metal enrichment history of the early Universe.

%remains one of the biggest uncertainties in , and more work is required before one can draw definite conclusions.
%whose physics is not included in the present physics.
 %This is further illustrated.

\begin{acknowledgements}
We are grateful to Volker Springel, Joop Schaye, and Claudio Dalla
Vecchia for letting us use their versions of \textsc{gadget} and their data
visualization and analysis tools. We thank Jun-Hwan Choi for discussions 
of the simulation set up. The authors would like to thank the referee for the 
constructive comments that significantly improve the quality of the manuscript. 
The simulations were performed with the Texas Advanced 
Computing Center (TACC) at The University of Texas at Austin for providing 
HPC resources under XSEDE allocation TG-AST160038 (PI's M. Jeon). The authors acknowledge 
the El Gato cluster at the University of Arizona, which is funded by the National 
Science Foundation through Grant No. 1228509. VB was supported by NSF
grant AST-1413501.
\end{acknowledgements}

\appendix
 \begin{figure*}[h]
\centering
  \includegraphics[width=170mm]{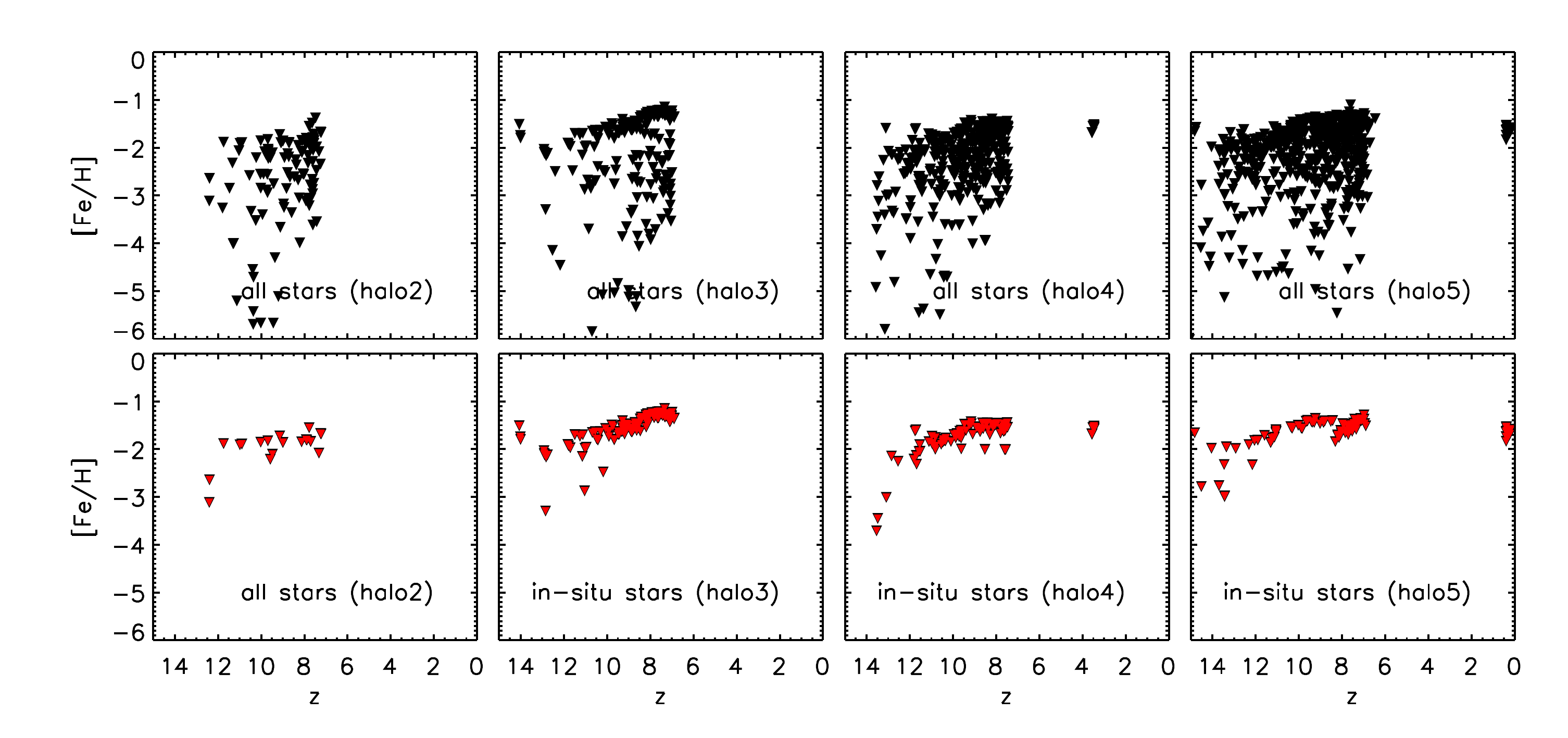}%{sedov_test.eps}
   \caption{The same as Figure~5, but for {\sc Halo2, Halo3, Halo4} and {\sc Halo5} from left to right. 
   Externally vs. internally formed stars. {\it Top panel:} stellar
   metallicity vs. formation time for {\it all} stars located within the virial radius of a halo at $z=0$.
   {\it Bottom panel:} subset of stars formed {\it in-situ} in the primary progenitor of a halo.}
\end{figure*}

\bibliographystyle{apj}
\bibliography{apj-jour,myrefs2}

\end{document}